%% file: mnras_template.tex
\documentclass[fleqn,usenatbib]{mnras}

\usepackage{newtxtext,newtxmath}

\usepackage[T1]{fontenc}

\DeclareRobustCommand{\VAN}[3]{#2}
\let\VANthebibliography\thebibliography
\def\thebibliography{\DeclareRobustCommand{\VAN}[3]{##3}\VANthebibliography}

\usepackage{graphicx}	
\usepackage{amsmath}	
\usepackage{physics}
\usepackage{siunitx}
\usepackage{multirow}
\usepackage{xcolor} 
\usepackage{subcaption}

\DeclareSIUnit \h {h}
\DeclareSIUnit \parsec {pc}
\DeclareSIUnit \pc {pc}
\DeclareSIUnit \cMpc {cMpc}
\DeclareSIUnit \year {yr}

\title[\texttt{PINION}]{\texttt{PINION}: Physics-informed neural network for accelerating radiative transfer simulations for cosmic reionization}

\author[D. Korber et al.]{
Damien Korber,$^{1, 2}$ \thanks{Corresponding author: Damien Korber \href{mailto:damien.korber@unige.ch}{damien.korber@unige.ch}}
Michele Bianco,$^{1}$
Emma Tolley$^{1}$
and Jean-Paul Kneib$^{1}$
\\
$^{1}$Institute of Physics, Laboratory of Astrophysics, École Polytechnique Fédérale de Lausanne (EPFL), 1290 Sauverny, Switzerland\\
$^{2}$Observatoire de Genève, Université de Genève, Chemin Pegasi 51, 1290 Versoix, Switzerland
}

\date{Accepted 2023 February 22. Received 2023 February 13; in original form 2022 August 26}

\pubyear{2022}

\begin{document}

\label{firstpage}
\pagerange{\pageref{firstpage}--\pageref{lastpage}}
\maketitle

\begin{abstract}

    With the advent of the Square Kilometre Array Observatory (SKAO), scientists will be able to directly observe the Epoch of Reionization by mapping the distribution of neutral hydrogen at different redshifts. 
    While physically motivated results can be simulated with radiative transfer codes, these simulations are computationally expensive and can not readily produce the required scale and resolution simultaneously. 
    Here we introduce the Physics-Informed neural Network for reIONization (\texttt{PINION}), which can accurately and swiftly predict the complete 4-D hydrogen fraction evolution from the smoothed gas and mass density fields from pre-computed N-body simulation. 
    We trained \texttt{PINION} on the \texttt{C$^2$-Ray} simulation outputs and a physics constraint on the reionization chemistry equation is enforced. With only five redshift snapshots,
    \texttt{PINION} can accurately predict the entire reionization history between $z=6$ and $12$. 
    We evaluate the accuracy of our predictions by analysing the dimensionless power spectra and morphology statistics estimations against \texttt{C$^2$-Ray} results. We show that while the network's predictions are in very good agreement with simulation to redshift $z>7$, the network's accuracy suffers for $z<7$. 
    We motivate how \texttt{PINION} performance could be improved using additional inputs and potentially generalized to large-scale simulations.

\end{abstract}




\section{Introduction}
\input{doc/introduction.tex}

\section{Reionization Simulations} \label{chap:reionization_simulations}
\input{doc/simulation.tex}

\section{Evolution of Reionization} \label{chap:eor}
\input{doc/reionization.tex}

\newpage
\section{Physics Informed Neural Network for Reionization} \label{chap:ccnn}
\input{doc/network}

\section{Results and Analysis} \label{chap:res_ana}
\input{doc/results_analysis_v2}

\section{Conclusions}

\label{cha:conclusion}
\texttt{PINION} successfully predicts the reionization history from RT simulation standard inputs with high accuracy. \texttt{PINION} could be applied to even larger volume simulations to create radiative transfer simulations of unprecedented volume and resolution. Additionally, the physics-informed $\mathcal{L}_\mathrm{ODE}$ drastically reduces the amount of data required to train the network, with almost identical performance between the LD and PFD scenarios.

\texttt{PINION} is relatively fast for a reionization simulation.
Training the network for each scenario takes about 1.5 GPU-hours and the prediction for each scenario takes less than 84 GPU-hours. However, the prediction is an already highly parallelizable task that can be further improved. The data I/O is a major bottleneck in network training and prediction, and optimization was not explored as part of this work. For example, the prediction for a $300^3$ volume is done in parallel on 1000 GPU nodes, each working on a smaller subvolume. In the current implementation, each node loads the entire dataset and then predicts all the pixels in the subvolume in approximately five minutes. This could easily be reduced to less than two minutes by only reading the subvolume.

While \texttt{PINION} is already a viable candidate for reproducing the \texttt{C$^2$-Ray} simulations, there is still room for improvement. First, no hyperparameter optimization was performed for the training, so tweaks to the optimization algorithm or network structure may improve performance. Second, the smoothed field likely introduces strong systematic biases into the network's predictions. Starting from a more accurate proxy for the ionization front or the photoionization rate would result in more accurate predictions. Using an approximate but more physically-motivated RT simulation such as \texttt{Grizzly} as an input field may drastically improve performance while keeping computation times low. Finally, the data bias could be reduced by modifying the data loss to account for imbalanced data.

\texttt{PINION} is a promising candidate for almost unsupervised learning. The network requires very little input data to train, and the ODE loss only requires a realistic estimation of the photoionization rate for the calculation of \autoref{eq:ODE_loss}, which in this work we took instead the $\Gamma$ from \texttt{C$^2$-Ray} simulations. Moreover, we demonstrated that by providing a simplistic approach in \autoref{eq:mfp_photons} for calculating the smoothed field, the network can retrieve information on the redshift evolution observed by the photoionization rate maps, \autoref{fig:evolution_of_irate}. In principle, any mean free path model for the UV radiation could be used as a proxy in the smoothed field.

\section*{Acknowledgements}
MB acknowledges the financial support from the SNSF under the Sinergia Astrosignals grant (CRSII5\_193826). 
We acknowledge access to Piz Daint at the Swiss National Supercomputing Centre, Switzerland under the SKA's share with the project ID sk02.
This work has been done in partnership of the SKACH consortium through funding by SERI.
The authors would like to thank Prof. Ilian Iliev for valuable discussions and comments and for providing the \texttt{C$^2$-Ray} simulation outputs employed in this work. The authors would like to thank the referee for their valuable comments and suggestions that have improved the quality of this manuscript.

\section*{Data Availability}
The data underlying this article is available upon request, and can also be re-generated from scratch using the publicly available \texttt{C$^2$Ray}, \texttt{CUBEP$^3$M} and \texttt{Tools21cm} codes. The \texttt{PINION} code (\href{https://ascl.net/2209.008}{ascl:2209.008}) and its trained network weights are available on the EPFL radio astronomy \texttt{GitHub} page: \url{https://github.com/epfl-radio-astro/PINION}.
 


\bibliographystyle{mnras}
\bibliography{references.bib, ref_michele.bib} 




\appendix
\section{Validity of the data pre-processing}\label{app:valididy}

As described in \S~\ref{sec:prop_mask}, to encode additional information about the expanding behaviour of the ionization rate map in the PINION input data, we pre-processed the source field to create the smoothed source distribution field.
This field is a very rough approximation of this behaviour and should not be used outside \texttt{PINION}'s context, as the CNN component of \texttt{PINION} is needed to correct for accurate physical behavior. 

\begin{figure}
    \centering
    \includegraphics[width=0.45\textwidth]{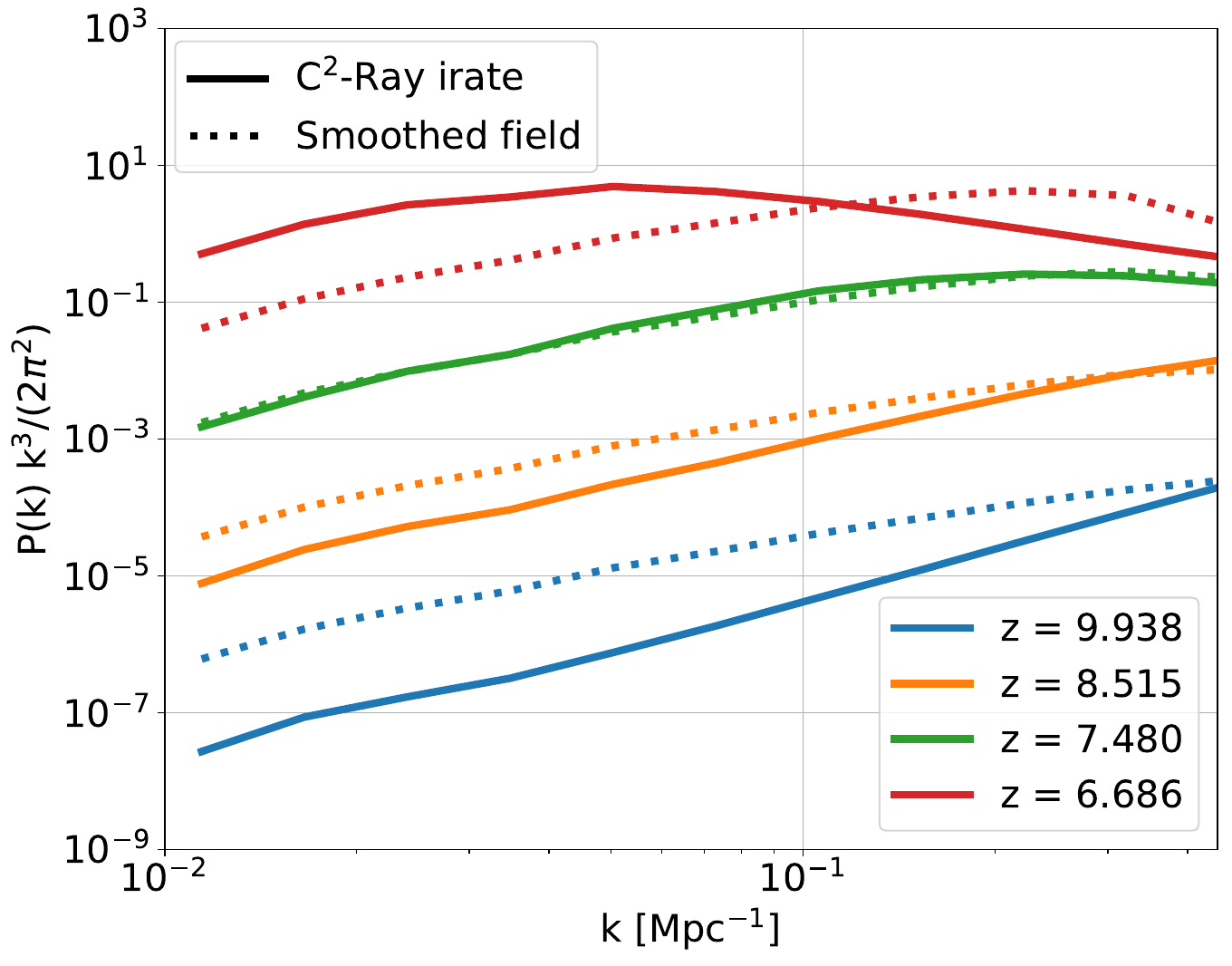}
    \vskip-4mm
    \caption{Dimensionless power spectrum of the photoionization rate and the smoothed source distribution field. As these are two different quantities (different scales and different units), these two field are transformed into dimensionless quantities before computing the power spectrum.}
    \label{fig:irate_ps}
\end{figure}

However, large differences between this smoothed field and the ionization rate map might introduce issues in the reconstitution of the $x_\text{HII}$ map, as seen in \S~\ref{sec:interp}.
We highlight the behavioural differences between the ionization rate and the smoothed field in \autoref{fig:irate_ps}.
This figure shows the dimensionless power spectrum for the dimensionless quantity $\hat{x} = x / \bar{x} - 1$, where $x$ correspond to the photoionization rate field and the smoothed source distribution field respectively. This quantity compares the scale dependence of structures caused by difference in the two fields, and any divergence can be fixed by adjusting a scaling factor. This figure shows that the dimensionless power spectrum is well reproduced for redshifts around $z\approx7.5$. However, for earlier redshifts ($z \gtrsim 8.5$) the smoothed field over-predicts large-scale structure $k<10^{-1}\,\mathrm{Mpc}^{-1}$, while it is in relative good agreement with the small scale. For later redshifts ($z \lesssim 7$), the smoothed field  under-predicts the small scale structures, while over-predicting the large-scale structures.

\section{Modifying the smoothed field mean free path}\label{app:f_mfp}
\begin{figure*}
    \centering
    \makebox[\textwidth][c]{\includegraphics[width=1\textwidth]{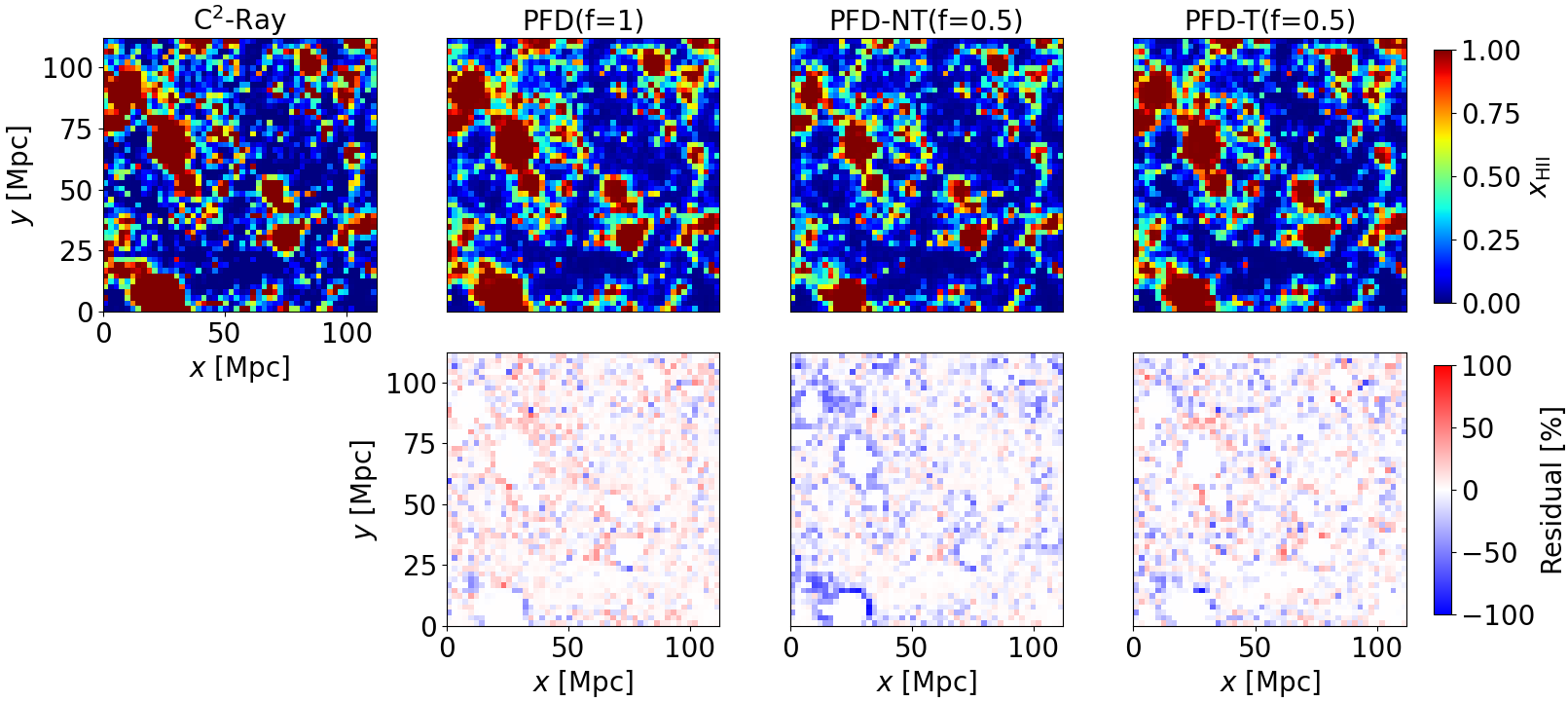}}
    \caption{Visualization of a small cutout of the ionized hydrogen map at redshift $z=\num{7.305}$. The first two columns are the same as in \autoref{fig:visualization_zoom}, $f=1$ indicating no modification to the mean free path. The third and fourth columns correspond to the prediction using the $f=0.5$ smoothed field without modifying the training and with a retrained network. Axes and colour bars are shared for readability.}
    \label{fig:visualization_zoom_f}
\end{figure*}

\begin{figure*}
    \centering
    \makebox[\textwidth][c]{\includegraphics[width=1\textwidth]{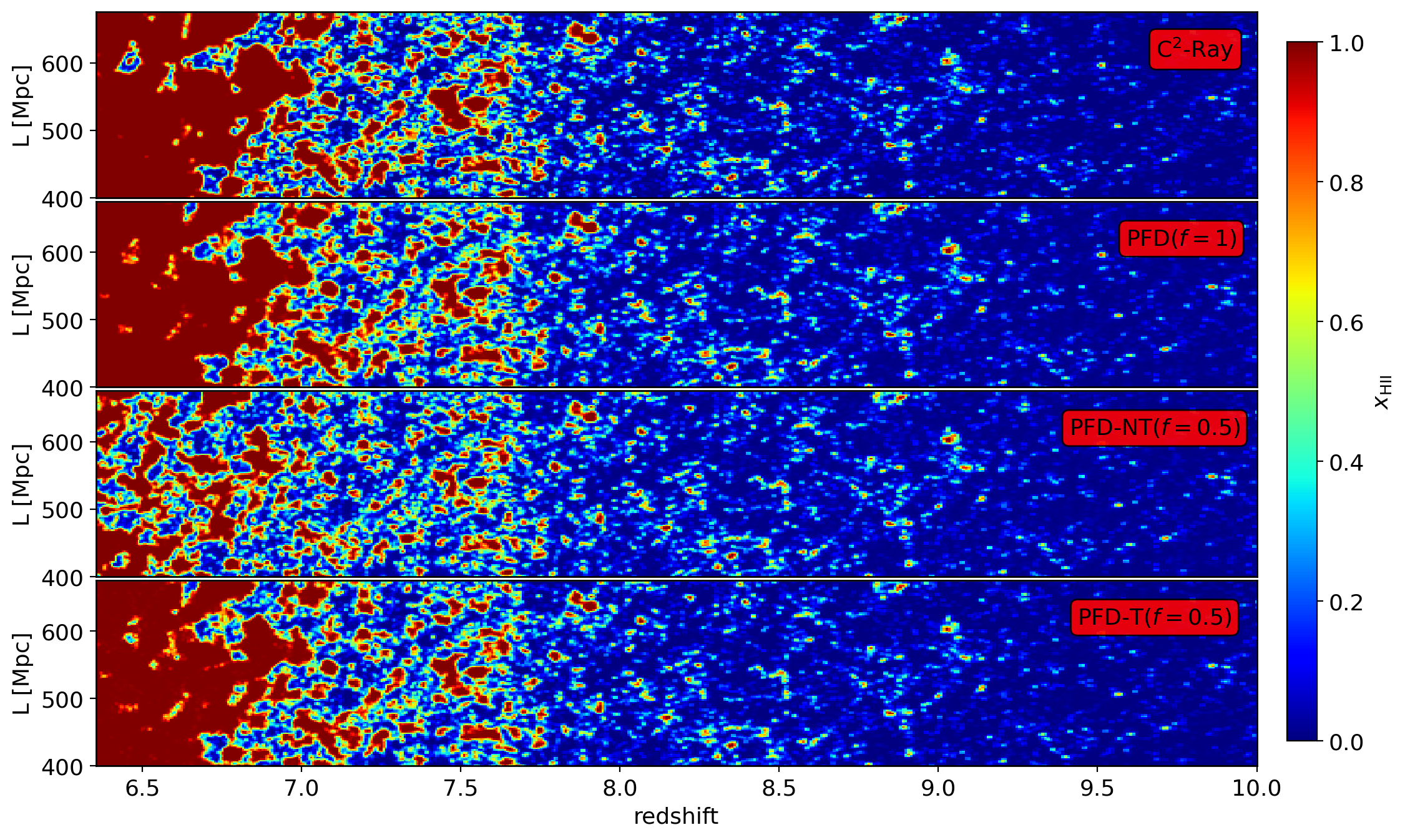}}
    \vskip+2mm
    \caption{Slices through the redshift axis of the ionized hydrogen light-cone. The colour bar indicates the ionization fraction. For readability, we cut at redshift $z=10$ and position $L\in[400, 600]\,\rm{Mpc}$.}
    \label{fig:light_cone_f}
\end{figure*}

\begin{figure}
    \centering
    \includegraphics[width=.5\textwidth]{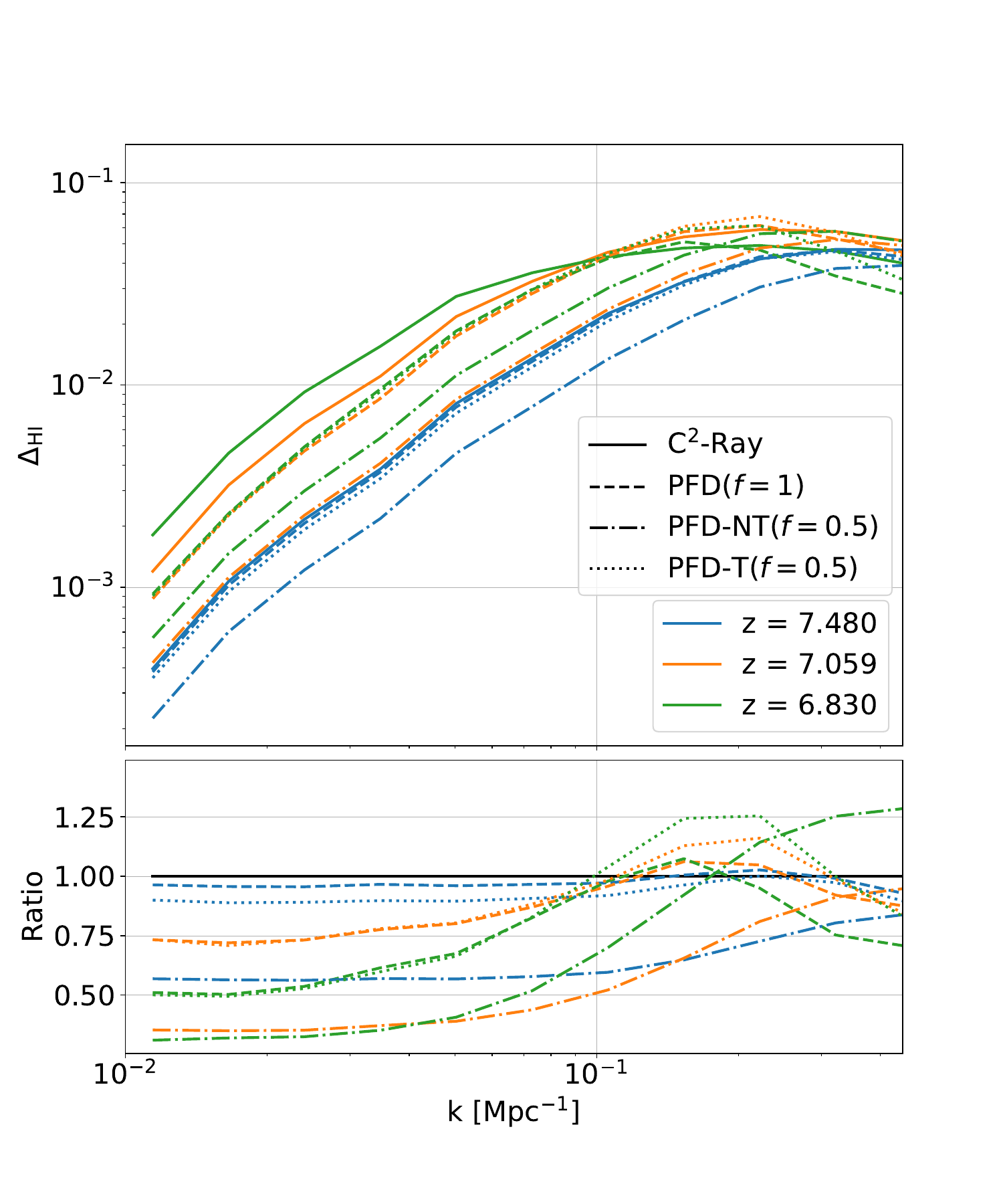}
    \vskip-10mm
    \caption{\textit{Top panel}: Dimensionless power spectrum of $x_\mathrm{HI}$. Solid lines are \texttt{C$^2$-Ray}, dashed lines are PFD ($f=1$) scenario, dash-dot lines are PFD-NT ($f=0.5$) scenario and dotted lines are PFD-T ($f=0.5$) scenario. These lines are chosen for the same redshifts. The blue line corresponds to approximately $\bar{x}_\mathrm{HI} = 0.7$, the orange to $\bar{x}_\mathrm{HI} = 0.5$ and the green to $\bar{x}_\mathrm{HI} = 0.3$. \textit{Bottom panel}: Ratio plot for the above dimensionless power spectrum. The ratio is obtained by dividing the predicted power spectrum with \texttt{C$^2$-Ray}'s power spectrum.}
    \label{fig:ps_f}
\end{figure}

To evaluate the impact of the mean free path for the smoothed fields, we recalculated the smoothed field with the mean free path reduced by a multiplicative factor $f = 0.5$, physically equivalent to the ionizing photons travelling half the distance at a given redshift.
The full analysis described in the paper was run under this modification, with and without network retraining (T and NT corresponding to \textit{Trained} and \textit{Not-Trained}, respectively). The resulting figures for redshift $z=7.305$ are shown in \autoref{fig:visualization_zoom_f}, \autoref{fig:light_cone_f} and \autoref{fig:ps_f}. The mean fraction of ionized hydrogen over the whole cube at this redshift is $\bar{x}_\mathrm{HII} = \num{0.389}$ for \texttt{C$^2$-Ray}, $\bar{x}_\mathrm{HII} = \num{0.398}$ for PFD($f=1$), $\bar{x}_\mathrm{HII} = \num{0.308}$ for PFD-NT($f=0.5$) and $\bar{x}_\mathrm{HII} = \num{0.370}$ for PFD-T($f=0.5$). These plots show that the network outputs are agnostic to a change in the mean free path scale as long as the network is retrained using the modified smoothed field. Indeed, the PSD ($f=1$) case, presented in \S\ref{sec:network_output}, and the corresponding retrained case results, PFD-T ($f=0.5$), are very similar in these figures, indicating that the network can retrieve a reasonable ionization map as long as it is trained with some smoothed pre-processed field. For the predicted results without proper training, PFD-NT ($f=0.5$), the reionization is, as expected, generally slower. The opposite behaviour was also observed for a factor $f=2$ but is omitted for brevity. This indicates the importance of proper training on the pre-processed inputs for \texttt{PINION} and the robustness of PINION to particular choices of pre-processing strategies.

\section{Mass-averaged analysis}\label{app:massVSvol}
\begin{figure}
    \centering
    \includegraphics[width=.37\textwidth]{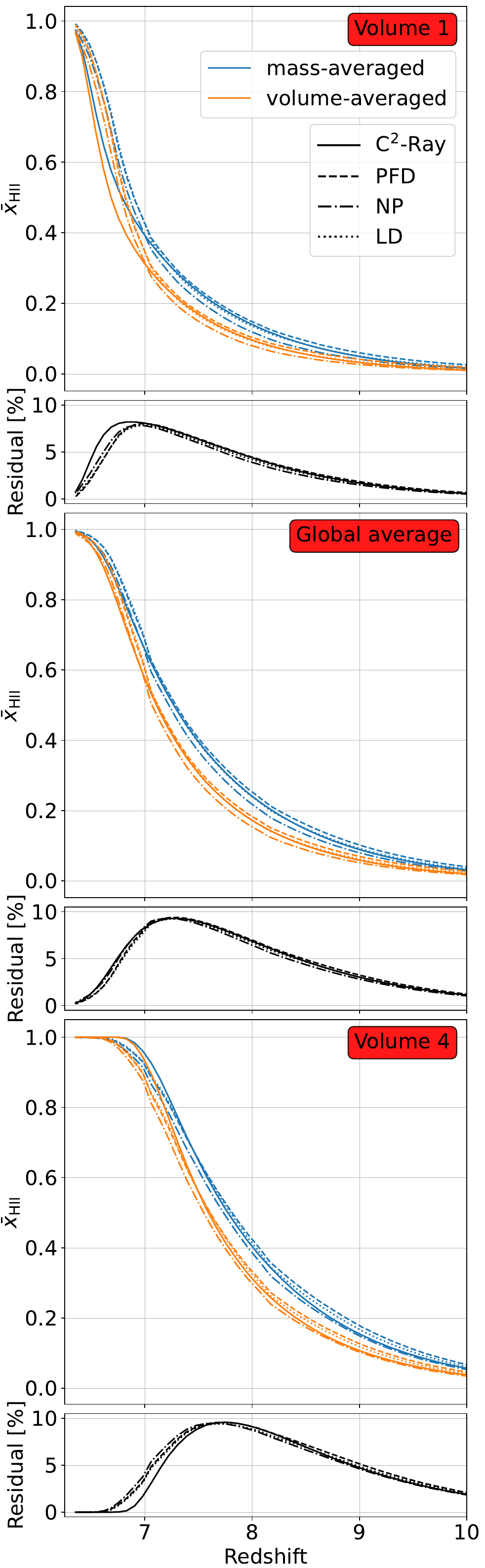}
    \caption{Comparison between the mass-averaged and volume-averaged ionization fraction for Vol. 1, 4 and global average, as defined in \autoref{fig:xhii_evo}. Each volume is separated into two panels, the top one showing the evolution of $\bar{x}_\mathrm{HII}$ over redshift the residual between the two for each scenario.}
    \label{fig:vol_v_mass}
\end{figure}

To complement the volume-averaged quantities analysis from \S~\ref{chap:res_ana}, we compared the main results from \autoref{fig:xhii_evo} with their mass-averaged counterparts. The mass-averaged ionization fraction is defined as
\begin{equation}
\bar{x}_\mathrm{HII}^\mathrm{m} = 1-\frac{\sum_i (1-x_{\mathrm{HII}, i})(1+\delta_i)}{\sum_i (1+\delta_i)}
    \label{eq:vol2mass}
\end{equation}
where $x_{\mathrm{HII}, i}$ is the volume-averaged ionization fraction and $(1+\delta_i)$ is the gas overdensity. The summation term $i$ is computed on the entire simulated volume. The comparison between the mass and volume-averaged quantities is shown in \autoref{fig:vol_v_mass}. One can notice that the general behaviour of large volume simulation, where the mass-averaged tend to ionize earlier than the volume-averaged counterparts \citep{iliev_simulating_2014}.

\bsp	
\label{lastpage}
\end{document}

%% file: doc/introduction.tex
Immediately after the Epoch of Recombination, no significant radiation sources existed in the Universe. This period is called the Dark Ages. However, the evolution of the matter density field led to the generation of collapsed structures, leading to the formation of the first stars and galaxies. At $z \sim 30$, these first luminous sources started to emit ultraviolet light, which successively ionized the intergalactic medium (IGM) \citep{Furlanetto2006, Ferrara2014}. Indirect constraints such as high-redshift quasar spectra \citep{Fan2006, McGreer2011, McGreer2014}, the decline of Lyman $\alpha$ emitting galaxies \citep{Schenker2011, Stark2011, Pentericci2014, Tilvi2014} suggest that reionization was complete for $z<6$. This change from a dark universe filled with neutral hydrogen to the fully ionized IGM is known as the Epoch of Reionization (EoR) \citep{gorbunov_introduction_2011, DAYAL20181}. The study of the EoR is essential for cosmology. The evolution of reionization is driven by the matter-energy content and geometry of the Universe, and the distribution of ionized regions reflects how galaxies and black holes formed.

The presence of neutral hydrogen during the EoR will be evident by measuring the emerging brightness temperature from a gas cloud at a given redshift against the CMB temperature \citep{zaroubi_epoch_2013}. Modern telescopes such as LOFAR\footnote{\url{http://lofar.org}} are able to measure the global statistics of the 21-cm signal \citep{Paciga2013, Yatawatta2013, Parsons2014, Jelic2014InitLofar, Jacobs2015, Patil2017, Mertens2020, Trott2020DeepObservations}. However, during reionization the signal fluctuations are expected to be non-Gaussian \citep{Mondal2017, Shimabukuro2017, Majumdar2018, Hutter2019, Gorce2019, Watkinson2022Bispectr} and therefore can not be fully depicted by the power spectra only. Next-generation telescopes such as the SKA\footnote{\url{https://skatelescope.org}} will be able not only to detect the 21-cm signal but also to produce maps of its fluctuations on the sky \citep{Mellema2013ReionizationArray, iliev_simulating_2014, Dillon2015mapmaking, Wyithe2015ImagingSKA, Koopmans2015TheArray, SKA_RedBook2020}. These observations at multiple  frequencies will allow us to directly measure the neutral hydrogen density distribution throughout the Universe evolution. The expected precision of SKA-Low will allow a groundbreaking comparison between observations and simulations. Future high-fidelity observations of the EoR demand equivalently high-fidelity simulations. A physically motivated simulation of reionization needs to include three main components: gravitational evolution of dark matter particles via N-Body simulations, gas dynamics via hydrodynamical simulations, and radiative transfer (RT) simulations to compute the propagation of photons and the evolution of the ionized atoms. Although it is possible to do \textit{on-the-fly} simulations that combine these three steps \citep{Rosdahl2013Ramses, Ocvirk2016coda, Ocvirk2020codaii, Ocvirk2021codaiii, Lewis2022}, these simulations are very computationally expensive as they require small time steps and large number of simulated particles \citep{Abel1999, Bolton2004, Shapiro2004}. It is also possible to \textit{post-process} the RT step onto existing N-body and hydrodynamical simulations in order to lighten its computational footprint. The state-of-the art post-process RT simulation is the \texttt{C$^2$-Ray} code \citep{mellema_c2-ray_2006}. This method uses a 3-D ray tracing method that relaxes the constraints on the time step size. While \texttt{C$^2$-Ray} can produce high-accuracy EoR simulations at large scale, is extremely computationally expensive to run, requiring several millions of core-hours to run across thousands of nodes.

Another post-process RT code is \texttt{Grizzly}  \citep{ghara_prediction_2018}, which assumes spherical density profiles around each source, thus simplifying the problem to a 1-D RT process and speeding up the computation by several orders of magnitude. There is no simulation which is able to produce RT on large volumes while still accounting for small scale physics.

Deep neural networks are well-known as universal function approximators \citep{hornik_approximation_1991}, and have been successfully used to accelerate and improve simulations of physical processes in many domains \citep{doi:10.1073/pnas.1821458116, LIU2021109152}. Deep learning has previously been applied in the context of reionization simulations. In particular, \citep{chardin_deep_2019} obtained some encouraging results for the use of neural networks in predicting the 3-D ionization map at $z = 6$.

Despite their success, neural networks suffer from interpretability and generalization problems \citep{7298640, 10.1145/2939672.2939778}. While physical phenomenons are often described by some theoretical or empirical model, neural networks attempt to learn these laws from statistically small and biased data. However, in recent years, a new class of neural networks with physics constrains has emerged. These networks are called \textit{Physics-Informed Neural Networks} (PINNs) and encapsulate additional physics laws into the behaviour of the network \citep{karniadakis_physics-informed_2021}.

In this paper, we present \texttt{PINION}, a physics-informed neural network which can accurately predict the reionization map at any redshift and volume. This paper is organized as follows. In \S~\ref{chap:reionization_simulations} we present the numerical codes employed to simulate the reionization process and create the network training dataset. In \S~\ref{chap:eor} we present the model used to study the evolution of ionization during the EoR and data pre-processing by smoothing the source distribution field to encapsulate the expanding behaviour of the ionising front. Then in \S~\ref{chap:ccnn} we present \texttt{PINION}, the neural network that we developed for this project, with its physics constraint. \S~\ref{chap:res_ana} gives the results and their analysis, and \S~\ref{cha:conclusion} concludes with a summary of the results and the future prospects for this project.

%% file: doc/simulation.tex
The simulation used in this paper is briefly described here below, but we refer to the reader the recent works \citep[e.g.][]{dixon_large-scale_2016, ross_simulating_2017, bianco_impact_2021} for further details.
The cosmological parameters used in the simulation are based on the WMPA three-year data observation and the results are consistent for a $\Lambda$CDM cosmology with the following parameters, $\Omega_{\Lambda}=\num{0.724}$, $\Omega_{m}=\num{0.276}$, $\Omega_{b}=\num{0.454}$, $H_0=\SI{70.1}{\km\per\s\per\mega\pc}$, $\sigma_8=\num{0.784}$ and $n_s=\num{0.946}$ \citep{spergel_three-year_2007}. The \texttt{CUBEP}$^\texttt{3}$\texttt{M} code \citep{Harnois-Deraps2013} is employed for the N-body simulation and it is run on a box of physical size $714\,\rm{Mpc}$ containing $6912^3$ particles, with a spatial resolution of $5.17\,\rm{kpc}$ and particles mass of $5.12\times10^3\,M_{\odot}$. The initial conditions are generated using the Zel’dovich approximation, while the power spectrum of the linear fluctuations is given by the \texttt{CAMB} code \citep{Lewis2000}. The simulation then starts at redshift $z = 150$, which gives enough time to significantly reduce the non-linear decaying modes \citep{Crocce2006}. An on-the-fly spherical over-density halo finder algorithm \citep{Harnois-Deraps2013, Watson2013}
with over-density parameter $\Delta=130$
creates a halo list catalogues at each redshift step, while the remaining particles are considered to be part of the IGM.
In this study, we only consider  halos with mass $M_{h}\geq 10^9\,M_{\odot}$ as these sources are not affected by radiative feedback and are considered to be the primary driver of reionization \citep{Iliev2012, dixon_large-scale_2016}. 

The N-body particle list and the halo catalogues are then interpolated into a $300^3$ grid with a Smoothed-Particle-Hydrodynamic-like method \citep{Shapiro1996, Mao2012}, with a corresponding cell size of $2.381\,\rm{Mpc}$. This dark matter density field and cumulative source mass field are used as inputs to the \texttt{C$^2$-Ray} code, which computes the post-processed RT to generate the photoionization rates and ionization fractions. A fixed time step $\Delta t = 11.54 \rm{Myr}$ is considered for redshifts $z=$ 6 -- 50. We refer the reader to, e.g. \citep{dixon_large-scale_2016, majumdar_effects_2016, bianco_impact_2021} for a more general overview of the RT method and N-Body method employed in this paper. In our case, snapshots of \texttt{C$^2$-Ray} for the redshifts range  $z=$ 6 -- 12 were kept for training the network, which corresponds to the time when at least one source is present at each pixel of the smoothed grid and therefore where the reionization process is rapidly evolving.

The  matter density field and  source mass field used by \texttt{C$^2$-Ray} are also provided as inputs to the neural network.

%% file: doc/reionization.tex
In the EoR we are interested in studying how the neutral hydrogen density $n_\mathrm{HI}$ is converted into the ionized hydrogen density $n_\mathrm{HII}$. Reionization is driven by two different processes: {\bf ionization}, by which high-energy photons ionize neutral hydrogen, and {\bf recombination}, by which ionized hydrogen recombines with the local electron density $n_\mathrm{e}$ into neutral hydrogen. The evolution of ionized hydrogen is given by \citep{choudhury_analytical_2009, Pritchard2012}:

\begin{equation}
    \frac{dn_\mathrm{HII}}{dt} = n_\mathrm{HI}\Gamma - \mathcal{C}\alpha_\mathrm{B} n_\mathrm{HII} n_\mathrm{e}
    \label{eq:first_ODE}
\end{equation}
where $\Gamma$ is the reionization rate, $\mathcal{C}$ is the clamping factor which accounts for the inhomogeneity in the simulation, and $\alpha_\mathrm{B}\equiv\alpha_\mathrm{B}(T)=2.59\times10^{-13} (T/10^4\,\mathrm{K})^{-0.7} \mathrm{cm^3\,s^{-1}}$ is the temperature dependent case B recombination coefficient \citep{Furlanetto2006}. In this work, the clumping factor is set to $\mathcal{C} = 1$. This scenario corresponds to the case where sub-grid inhomogeneities in the IGM are ignored \citep{Mao2019clump, bianco_impact_2021}. Moreover, we set the temperature to $T=10^4\,\mathrm{K}$, which assumes that gas in halo is able to radiatively cool through hydrogen and helium atomic lines \citep{Wise2014}. For a fixed density of hydrogen ${n}_\mathrm{H} \equiv n_\mathrm{HI} + n_\mathrm{HII}$, we can define the neutral hydrogen fraction $x_\mathrm{HI} \equiv n_\mathrm{HI} / n_\mathrm{H}$ and ionized hydrogen fraction $x_\mathrm{HII} \equiv n_\mathrm{HII} / n_\mathrm{H}$. For simplicity, we assume pure hydrogen gas, $n_\mathrm{e} \sim n_\mathrm{HII}$. We can then rewrite equation \eqref{eq:first_ODE} as:

\begin{equation}
    \frac{dx_\mathrm{HII}}{dt} = (1- x_\mathrm{HII})\Gamma - \mathcal{C}\alpha_\mathrm{B} n_\mathrm{H} x^2_\mathrm{HII}
    \label{eq:simplified_ODE}
\end{equation}

Note that in the following of this paper, we will refer to $x_{\mathrm{HII}}$ as the volume-average fraction of ionized hydrogen.

\begin{figure*}
    \centering
    \makebox[\textwidth][c]{\includegraphics[width=\textwidth]{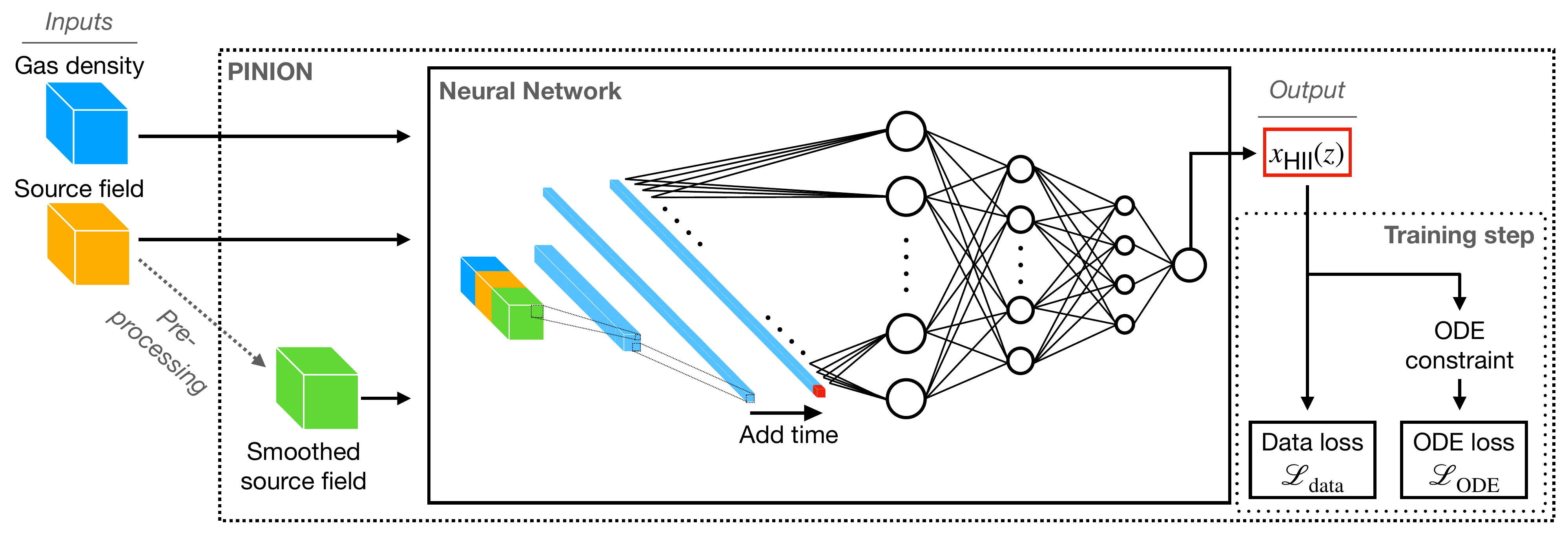}}
    \vskip-4mm
    \caption{Diagram of \texttt{PINION}'s pipeline. The input data consist in two input subvolumes, the gas density and the source field that were produced by the N-Body simulation. The source field is then processed to obtain a smoothed field. The network outputs one value for the subvolume central pixel $x_\text{HI}$. During the training steps, the output is then compared to the data to obtain the data loss $\mathcal{L}_\text{data}$ and the physics constraint to obtain the the ODE loss $\mathcal{L}_\text{ODE}$.}
    \label{fig:pipeline}
\end{figure*}

\begin{figure*}
    \centering
    \begin{subfigure}{\textwidth}
        \includegraphics[width=\textwidth]{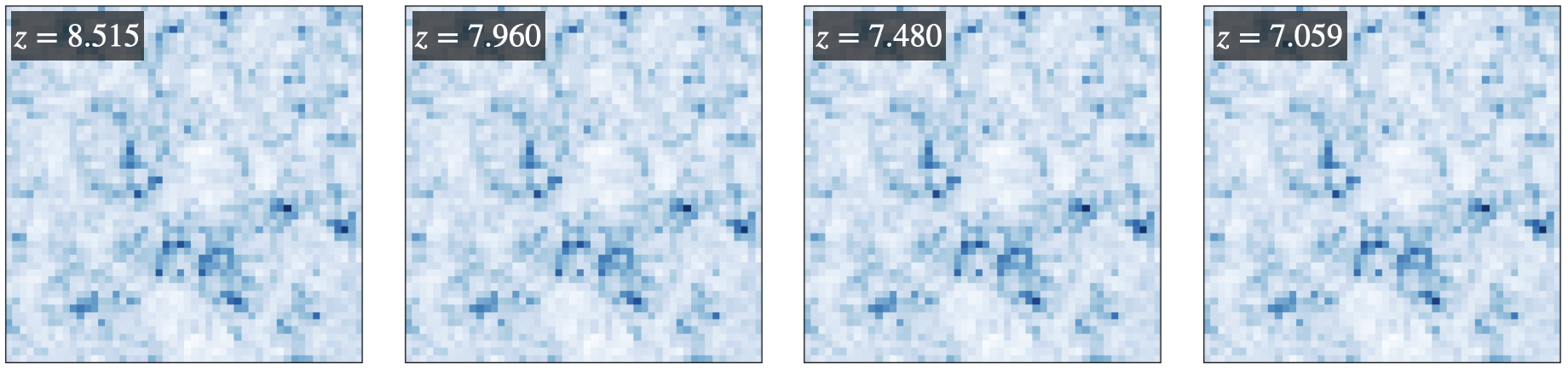}\vskip-2mm
        \caption{Evolution of the gas density: light blue are under-dense regions and dark blue are over-dense regions.}
        \label{fig:evolution_of_rho}
    \end{subfigure}
    ~
    \begin{subfigure}{\textwidth}
        \includegraphics[width=\textwidth]{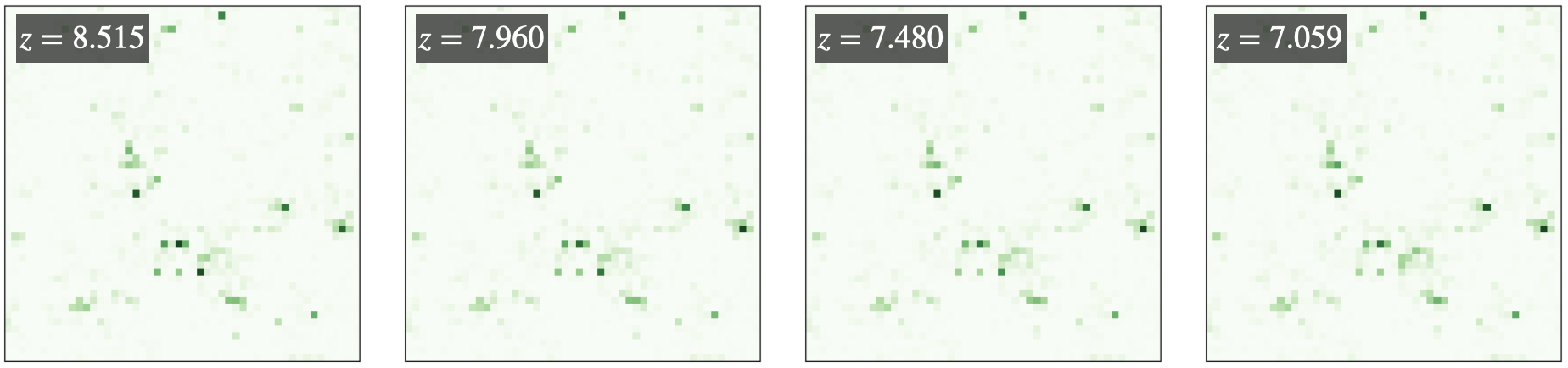}\vskip-2mm
        \caption{Evolution of the mass of sources: light green are regions with few particles, and dark green are the sources.}
        \label{fig:evolution_of_nsrc}
    \end{subfigure}
    ~
    \begin{subfigure}{\textwidth}
        \includegraphics[width=\textwidth]{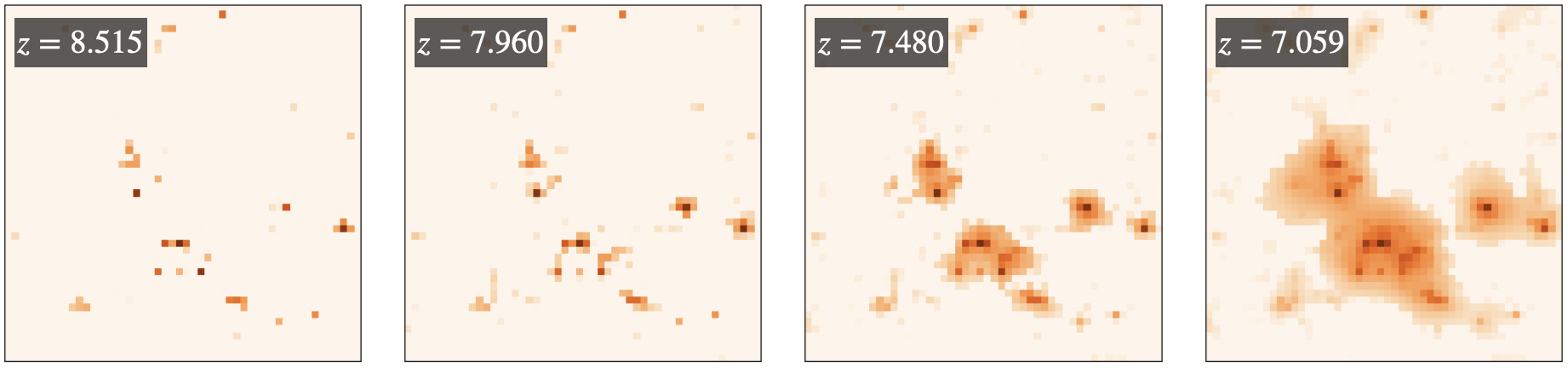}\vskip-2mm
        \caption{Evolution of the photoionization rate: light orange are regions which receives less ionizing photons and dark orange are the ionizing sources.}
        \label{fig:evolution_of_irate}
    \end{subfigure}
    ~
    \begin{subfigure}{\textwidth}
        \includegraphics[width=\textwidth]{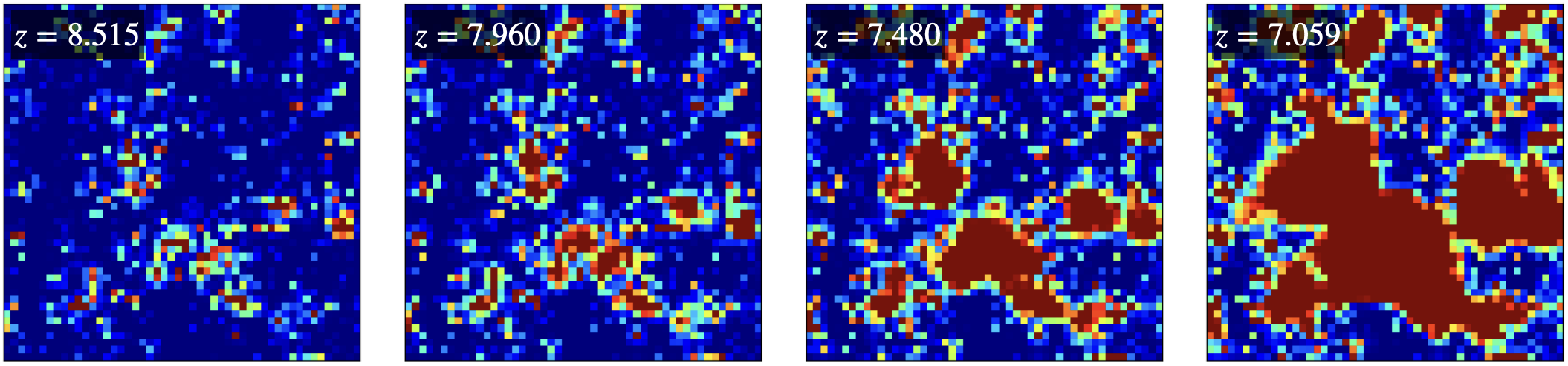}\vskip-2mm
        \caption{Evolution of the ionized fraction: red are ionized regions and dark blue are neutral regions.}
        \label{fig:evolution_of_xhii}
    \end{subfigure}
    \caption{Evolution of the gas density, mass of sources, photoionization rate, and ionized fraction for four different redshifts. Note that the scale is different for every plot but the evolution shows that \autoref{fig:evolution_of_rho} and \ref{fig:evolution_of_nsrc} are not expanding like \autoref{fig:evolution_of_irate} and \ref{fig:evolution_of_xhii}.}
    \label{fig:evolutions_of_qte}
\end{figure*}


%% file: doc/network.tex
\label{sec:CCNN_description}
\texttt{PINION}\footnote{\url{https://github.com/epfl-radio-astro/PINION}} (Physics Informed Neural network for reIONization) is a 3-D CNN (Convolutional Neural Network) that reduces the information from two $7^3$ input cubes of gas density and mass of sources to predict $x_\mathrm{HII}$ at the cube centre for a single value of redshift. The input size corresponds to a volume of $(16.67\,\mathrm{Mpc})^3$.
We initially chose a size scale based on the physical hard limit imposed by Lyman limit system \citep{Shukla2016LLS}, corresponding to an input mesh-grid size of $49^3$. However, we did not observe significant improvement in the result for this larger subvolume, and decided to reduce the input size to reduce the training and predicting computation time. We considered periodic boundary conditions on the full dataset to be able to predict the border pixels. This is justified by the fact that the \texttt{C$^2$-Ray} simulation we are training on also uses periodic boundary conditions.


\subsection{\texttt{PINION} pipeline and architecture}
\texttt{PINION} has 2 input channels: the source mass field and the gas density. An additional third input is derived internally by smoothing the source distribution field, described in \S~\ref{sec:prop_mask}. These three fields are then internally given as inputs to the network, which then returns the ionized fraction $x_\text{HII}$ of the subvolume central pixel as the output. The network is composed of three convolutional layers with $3^3$ kernel and doubling number of feature maps: 64, 128, 256. The convolution results in 256 feature maps with channel size $1$, to which the normalized look-back time is concatenated. This tensor is then used as input to a fully connected network with three hidden layers of 64, 16, and 4 nodes, and a single output node. Each layer uses the PReLU activation function \citep{prelu} followed by a 3-D batch normalization and a 10\% dropout. The output node uses a sigmoid activation function to truncate the output values between 0 and 1. In total, this network has 1,130,252 trainable parameters. In \autoref{fig:pipeline}, we show a simplified overview of \texttt{PINION} pipeline, the pre-process step and the neural network architecture.

We use the Adam optimization algorithm \citep{Kingma2015AdamAM} during the network training to minimize our data loss function $\mathcal{L}_\mathrm{data}$. In \texttt{PINION}, the data loss is the adjusted $R^2$ loss, also referred to the \textit{coefficient of determination} in statistics. This quantity is defined as follows:
\begin{equation}
    \mathcal{L}_\mathrm{data} (x_{\rm true},\,x_{\rm pred}) = \frac{\sum(x_{\rm pred} - x_{\rm true})^2}{\sum (x_{\rm true} - \bar{x}_{\rm true})^2} \equiv 1-R^2
\end{equation}
where $x_{\mathrm{pred}}$ is the prediction from the network, $x_{\mathrm{true}}$ is the ground truth and $\bar{x}_{\mathrm{true}}$ is its mean value \citep{gillet_deep_2019}.

We train with the ODE loss as defined in \S~\ref{sec:physics_constraint}. The results from this network will be discussed in \S~\ref{cha:results} and show promising results.

\subsection{Data pre-processing} \label{sec:prop_mask}

During initial studies, we found that it was difficult for the network to learn the relationship between the input fields from the N-body simulation and the reionization across all redshifts. This is because, as shown in \autoref{fig:evolutions_of_qte}, while the gas density and source mass fields do not change drastically over the EoR, the reionization rate and fraction rapidly change as ionized regions progressively grow with time. Without any other constraints, the network must learn to predict extremely different outputs from very similar inputs across all redshifts.

To ameliorate this problem, we provide the network with the ionizing front propagation as an additional input. This approach was inspired by \citep{doi:10.1073/pnas.1821458116}, which used a neural network to predict cosmological structure formation starting from the Zel’dovich Approximation \citep{zel1970gravitational} and in our case this input should give a crude approximation of the photoionization rate. We can calculate the mean free path of ionizing photons at a given redshift $\lambda_{\nu_\mathrm{HI}}(z)$ using the approximated form of equation (23) in \citep{choudhury_analytical_2009}:
\begin{equation}
    \lambda_{\nu_\mathrm{HI}}(z) \approx \frac{c}{H(z)}\times 0.1\left(\frac{1+z}{4}\right)^{-2.55}
    \label{eq:mfp_photons}
\end{equation}
This equation is the empirical fit from simulations valid for redshifts $z>3$. While the source model in \citep{choudhury_analytical_2009} does not exactly correspond to our simulation, it provides a good starting point for the network. We can then approximate the size of the ionized bubbles using $\lambda_{\nu_\mathrm{HI}}(z)$, and use the smoothed sources mass field to infer the photoionization rate. The algorithm is the following:
\begin{enumerate}
    \item Calculate $\lambda_{\nu_\mathrm{HI}}(z)$ for a given redshift $z$.
    \item Construct a spherical kernel of size $\lambda_{\nu_\mathrm{HI}}(z)$ by converting $\lambda_{\nu_\mathrm{HI}}(z)$ from \si{\mega\parsec} to number of pixels given a resolution of \SI{2.381}{\mega\parsec} per pixel, setting every pixel to 0 when further than distance $\lambda_{\nu_\mathrm{HI}}(z)$ to the centre and scale the other pixel value between 0 and 1, where 1 is the centre pixel and 0 the borders.
    \item Finally, convolve cube of mass of sources with the spherical kernel.
\end{enumerate}
This algorithm creates a smoothed source distribution field that behaves approximately like the photoionization rate. 
Knowing that $\lambda_{\nu_\mathrm{HI}}(z)$ increases with redshift, the ionization front expand as well.

The purpose of this smoothed field is to provide an additional input to the network, and outside this machine learning context, it should not be interpreted as an estimate of the photoionization rate.


\subsection{Physics constraint}
\label{sec:physics_constraint}

One of the most popular ways to encode physics constraints in a deep neural network is through learning bias \citep{karniadakis_physics-informed_2021}.
We have incorporated this in our network by using the ODE from \autoref{eq:simplified_ODE} as an additional loss:
\begin{equation}
    \mathcal{L}_\mathrm{ODE} = \frac{dx_{\mathrm{HII}} }{dt} - (1-x_{ \mathrm{HII} }) \Gamma + \alpha_\mathrm{B} n_\mathrm{H} x_{ \mathrm{HII} }^2
    \label{eq:ODE_loss}
\end{equation}
This ODE depends on the time derivative of $x_\mathrm{HII}$. While automatic differentiation can be used to obtain the exact value of the derivative \citep{autodiff}, it is unfortunately not feasible for a network with millions of parameters. 

Instead, we encode the ODE loss as a finite-difference, using the Runge-Kutta 4 (RK4) method \citep{scherer_computational_2013} to calculate the derivative. RK4 requires an intermediate step to perform derivations, so we simply take a step size two times larger than the time increment of the data. Let $x_n \equiv x_{ \mathrm{HII} }(t_n)$ the ionization fraction at the $n$-th time step and $\Gamma_n \equiv \Gamma(t_n)$ the photoionization rate at the $n$-th time step. 
Then, defining $h_n = \Delta t_n + \Delta t_{n+1}$ as the new time step, where $\Delta t_n$ is the time between snapshots $n$ and $n+1$, and $D_n = \alpha_\mathrm{B} n_{\mathrm{H},n}$, we can define the RK4 terms as:
\begin{subequations}
    \begin{align}
        k_1 &= (1-x_n)\Gamma_n - D_n x_n^2 \label{eq:ode_rk4_k1}\\
        k_2 &= (1-x_n-\frac{h_n}{2}k_1)\Gamma_{n+1} - D_{n+1}(x_n + \frac{h_n}{2}k_1)^2 \label{eq:ode_rk4_k2}\\
        k_3 &= (1-x_n-\frac{h_n}{2}k_2)\Gamma_{n+1} - D_{n+1}(x_n + \frac{h_n}{2}k_2)^2 \label{eq:ode_rk4_k3}\\
        k_4 &= (1-x_n-h_nk_3)\Gamma_{n+2} - D_{n+2}(x_n + h_nk_3)^2 \label{eq:ode_rk4_k4}\\
        x_{n+2}^* &= x_n + \frac{h_n}{6}\left[ k_1 + 2k_2 + 2k_3 + k_4\right] \equiv x_n + \frac{h_n}{6}\mathcal{K} \label{eq:ode_rk4_solution}
    \end{align}
    \label{eq:ode_rk4} 
\end{subequations}

where $x_n$ is the predicted mean fraction of ionized hydrogen, obtained as the output of the neural network, $x^*_{n+2}$ is the step obtained by computing the evolution of the ODE from $x_n$ and $\Gamma_n$ is the true photoionization rate from \texttt{C$^2$-Ray}. The large time step makes it difficult to keep the solution stable. Indeed, at higher redshifts, most of the Universe is neutral, but at lower redshifts the ionization quickly increases. With large time steps, RK4 has no issues tracking the evolution of the ODE at earlier times, as changes are slow. However, at later times, the photoionization rate is rapidly changing and the RK4 solution might drastically increase. To fix this issue, the \autoref{eq:ode_rk4_solution} is clipped between 0 and 1 to guarantee that the solution does not diverge.
One can now write the physics loss as follows.
\begin{subequations}
    \begin{align}
        \Delta x &= x_{n+2} -  \mathrm{MinMax}\left( \abs{x_n + \frac{h_n}{6}\mathcal{K}}, 0, 1\right) \label{eq:physics_loss_derivative}\\
        \mathcal{L}_\mathrm{ODE} (x_{n}) &= \frac{1}{N}\sum_{n=0}^N\left(\frac{\Delta x}{h_n} - (1-x_n)\Gamma_n + D_n x_n^2\right)^2 \label{eq:physics_loss_MSE}
    \end{align}
    \label{eq:physics_loss}
\end{subequations}

\begin{figure}
    \centering
    \includegraphics[width=0.45\textwidth]{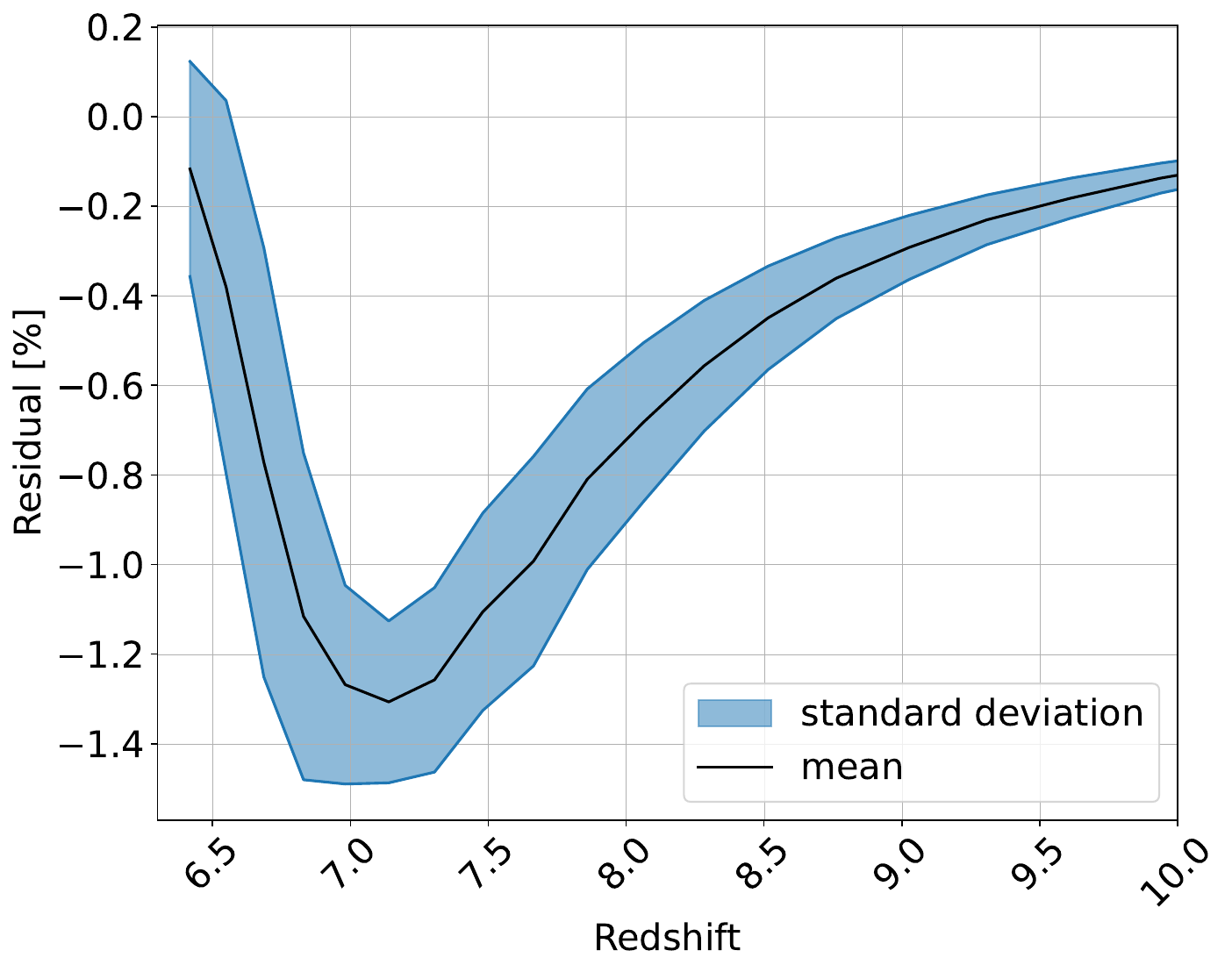}
    \vskip-4mm
    \caption{Residual between the RK4 ODE evolution of $x_\mathrm{HII}$ and the simulated $x_\mathrm{HII}$. This plot is obtained by averaging the evaluated ODE over 1,000 random subvolumes of size \SI{47.62}{\mega\pc}. The redshift is cut at $z=10$ to highlight the lower redshifts, where the residual is the biggest.}
    \label{fig:rk4_v_c2r}
\end{figure}

Where $\mathrm{MinMax}(x,0,1)$ ensures that $x$ is encapsulated between 0 and 1. 
In \autoref{fig:rk4_v_c2r}, we compute the residual to check that the RK4 loss accurately describe the behaviour of the \texttt{C$^2$-Ray} simulation.
We see good agreement between the data and the ODE as the standard deviation is within 2\% for all redshifts. When the reionization fraction is not rapidly evolving for  $z > 9$, we see excellent agreement within 0.5\%.

Although we use 3-D convolutional layers, we account for the time evolution constraint by using the batch dimension of the network input.
For a given spatial position of the input fields, its evolution for the 46 redshift snapshots is added as adjacent inputs along the batch dimension. The truth information is fed to the network in the same manner.
This way, we can use spatial convolution and study the evolution of the predicted quantity with the time evolution constraint from the ODE loss.  
This additional loss constrains the network's behaviour and improves performance.

\subsection{\texttt{PINION} and approximate RT solvers}

Compared to other fast approximations of RT, PINION offers a more physically motivated but less interpretable solution.

For example, radiative transfer codes such as \texttt{GRIZZLY} assume an uniform density profile around each isolated source \citep{ghara_prediction_2018}. This approximation reframes the reionization process as a 1D problem using rotational symmetries, which considerably simplifies the computations. Therefore, this approach computes the ionization profile around isolated sources, thus quantifying the radius of the spherical ionisation front as a function of time and optical depth \citep{Shapiro2005} and then redistributes photons in overlapping regions in a second step \citep{ghara_21_2015}. Recent works use 1D RT codes to study global statistics of the 21-cm signal for EoR and Cosmic Dawn \citep{datta_large_2022, kamran_redshifted_2022, ghara_constraining_2020, ghara_constraining_2021}.

\texttt{PINION} does not make any assumptions or approximations about the source or density field. However, PINION must infer the dynamics of RT during training, learning from a subvolume of size $(16.67\,\rm Mpc)^3 $ which we consider the \textit{region of influence}. We assume that this region contains all the nearby sources that contribute to the ionization state of the central pixel. During training, the ODE constraint \autoref{eq:physics_loss} enforces realistic behavior of the learned dynamics. However, as in all deep learning solutions, \texttt{PINION} results will have less interpretability compared to approximate 1D RT codes.

%% file: doc/results_analysis_v2.tex
\label{cha:results}
In order to characterize the impact of the ODE loss, we consider three different network training scenarios:
\begin{enumerate}
    \item NP (\textit{No Physics}): The training contains the entire evolution of pixels, but does not have a physics constraint:
$\mathcal{L}_\mathrm{total} = \mathcal{L}_\mathrm{data}$.
This scenario is noted by dashdot lines ($\cdot -$) in figures.

    \item PFD (\textit{Physics and Full Data}): The training uses the ODE loss and the data loss for the entire time evolution of pixels:
$\mathcal{L}_\mathrm{total} = \mathcal{L}_\mathrm{data} + \mathcal{L}_\mathrm{ODE} $.
 This scenario is noted by dashed lines ($--$) in figures.   

\item LD (\textit{Low Data}): The training uses the ODE loss, but the data loss is only evaluated at  5 different snapshots (limited) of the simulation instead of the full reionization history:
$\mathcal{L}_\mathrm{total} = \mathcal{L}_\mathrm{limited} + \mathcal{L}_\mathrm{ODE} $.
 This scenario is noted by dotted lines ($\cdot\cdot$) in figures.
\end{enumerate}

During the training, the three scenarios do not differ greatly in convergence time. Training each scenario took 1.5 GPU-hours on one node for 400 epochs. The training was performed on a node from CSCS's GPU cluster Piz Daint\footnote{\url{https://www.cscs.ch/computers/piz-daint/}}  that is equipped with NVIDIA Tesla P100 16GB GPU. In each scenario, only the network weights with the lowest validation loss is kept.

To predict the reionization fraction at a single pixel location, the network needs the surrounding $7^3$ pixel subvolume, corresponding to a volume of $(\SI{16.667}{\mega\pc})^3$, for the three internal input fields. These subvolumes are sliced out of the full cubes and are concatenated to form a higher dimensional tensor for the training dataset. For the PFD and NP scenarios, 4000 subvolumes over 46 redshifts are provided to the network (184,000 training subvolumes in total) with a batch size of 3680. For the LD training, 4000 subvolumes over 46 redshifts with a batch size of 3680 are also considered, but the data loss only had the ground truth for 5 fixed redshifts evenly spaced over the EoR evolution: ($z = \num{12.048}, \num{9.938}, \num{8.515}, \num{7.480}, \num{6.686}$) in the evolution, meaning that it compares the prediction to the truth data on 20,000 subvolumes instead of 184,000 for $\mathcal{L}_\mathrm{data}$. To validate the training, 500 validation subvolumes are considered, and the final network weights are selected from the training snapshot with the smallest total validation loss. Note that all the subvolumes are randomly chosen from the full dataset.

\subsection{Network output} \label{sec:network_output}
\begin{figure*}
    \centering
    \makebox[\textwidth][c]{\includegraphics[width=1\textwidth]{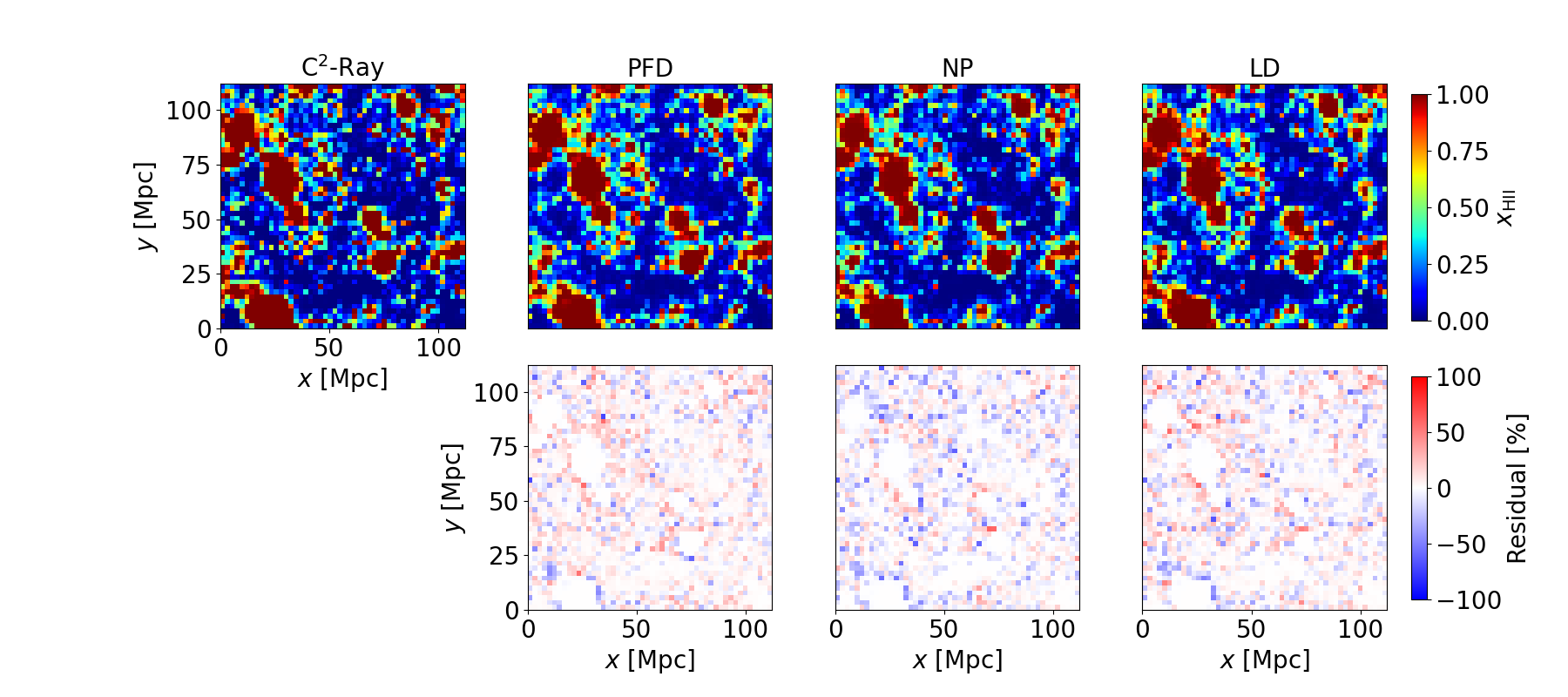}}
    \caption{Visualization of a small cutout of the ionized hydrogen map at redshift $z=\num{7.305}$ simulated by \texttt{C$^2$-Ray} and the three  \texttt{PINION} scenarios. The mean fraction of ionized hydrogen over the whole cube at this redshift is $\bar{x}_\mathrm{HII} = \num{0.389}$ for \texttt{C$^2$-Ray}, $\bar{x}_\mathrm{HII} = \num{0.398}$ for PFD, $\bar{x}_\mathrm{HII} = \num{0.364}$ for NP and $\bar{x}_\mathrm{HII} = \num{0.391}$ for LD.  The residual is computed as $R = x_\mathrm{HII}^\texttt{PINION} - x_\mathrm{HII}^\texttt{C$^2$-Ray}$. Axes and colour bars are shared for readability.}
    \label{fig:visualization_zoom}
\end{figure*}

The network results for each \texttt{PINION} training scenario are compared to the \texttt{C$^2$-Ray} ground truth in \autoref{fig:visualization_zoom} for a single redshift slice. In general, all the scenarios are able to accurately reproduce the ionization map. However, small differences can be observed at the borders of large and small ionized regions.

\begin{figure*}
    \centering
    \makebox[\textwidth][c]{\includegraphics[width=1\textwidth]{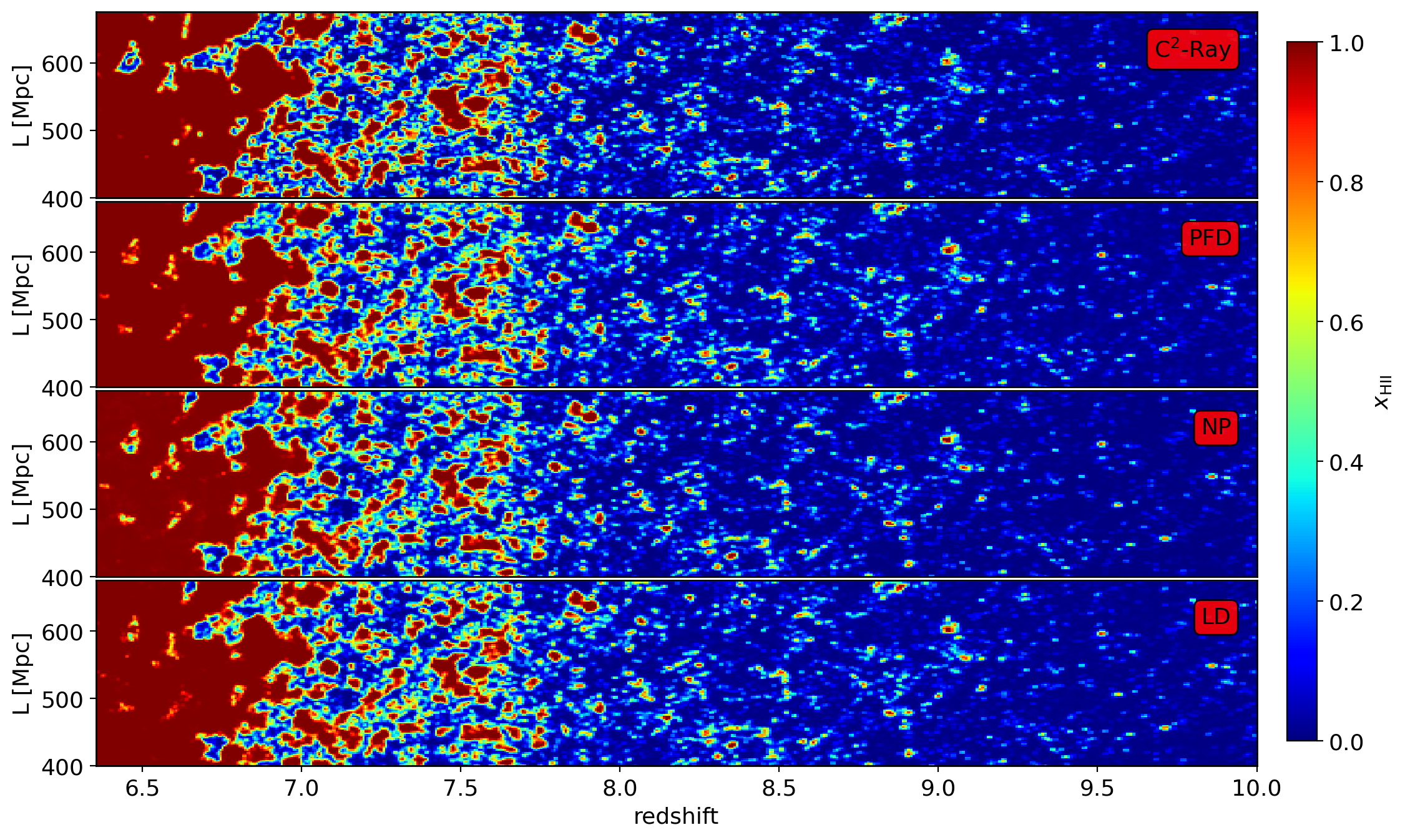}}
    \vskip+2mm
    \caption{Slices through the redshift axis of the ionized hydrogen light-cone, simulated by \texttt{C$^2$-Ray} and the three \texttt{PINION} scenarios. The colour bar indicates the ionization fraction. For readability, we cut at redshift $z=10$ and position $L\in[400, 600]\,\rm{Mpc}$.}
    \label{fig:light_cone}
\end{figure*}

The  $x_\mathrm{HII}$ redshift evolution in each scenario is shown next to \texttt{C$^2$-Ray} in \autoref{fig:light_cone}. The three scenarios evolve similarly and show good agreement with the ground truth. All the \texttt{PINION} scenarios are able to find the first ionizing bubbles ($z > 8$). See for example the bubble at $(z,\,L) = (9,\,610\,\rm{Mpc})$ and $(8.1,\,610\,\rm{Mpc})$ in \autoref{fig:light_cone}. However, for redshift $z<7$, we start to see large differences between \texttt{C$^2$-Ray} and the \texttt{PINION} predictions. For instance, the neutral island at $(z,\,L) = (6.4,\,600\,\rm{Mpc})$ or $(6.6,\,490\,\rm{Mpc})$ on the same figure. Between the different training scenarios, the NP scenario exhibits much more noise at low redshift ($z\approx6.5$) compared to the physics-constrained scenarios. Furthermore, we observe that neutral islands are smaller at this redshift in all predictions. Nevertheless, NP scenario appears to predict the most accurate neutral island size. 

\subsection{Evolution of the volume-averaged $\bar{x}_\mathrm{HII}$} \label{sec:evo_vol_av_x}

\begin{figure*}
    \centering
    \includegraphics[width=\textwidth]{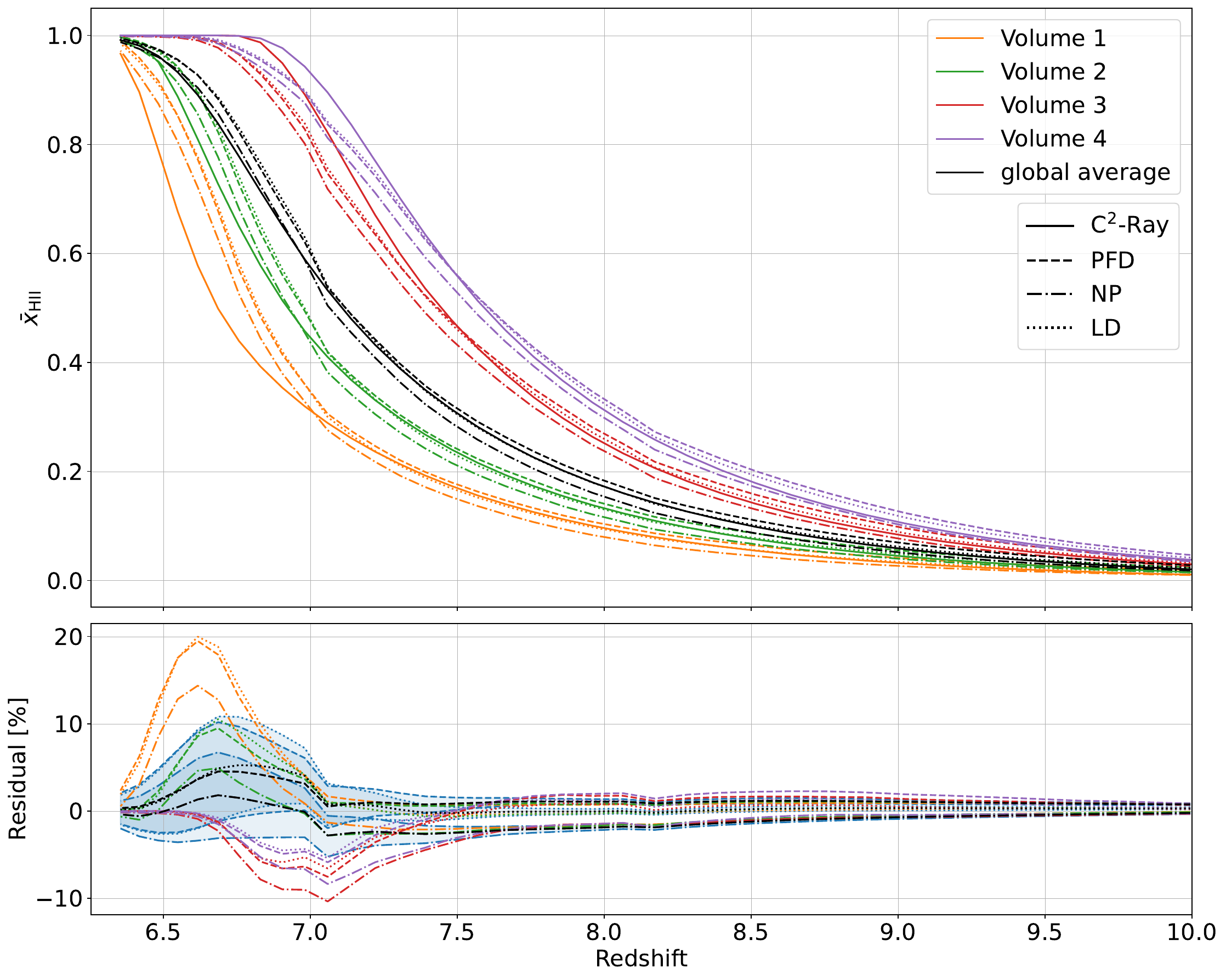}
    \vskip-4mm
    \caption{\textit{Top panel}: evolution of the mean fraction of ionized hydrogen for subvolumes of size \SI{47.61}{\mega\pc}. Four volumes with different end of ionization were hand-picked and are displayed with different colours. Black lines represent the global average for the whole volume. Solid lines are \texttt{C$^2$-Ray}, dashed lines are PFD training, dashdot lines are NP training and dotted lines are LD training. The x-axis is common to both plot and was cropped at redshift $z=10$ for readability. Some characteristic redshifts can be found in \autoref{tab:xhii_evo}. \textit{Bottom panel}: The residual between the \texttt{PINION} scenarios and the \texttt{C$^2$-Ray} simulation for the above evolutions are displayed with the same colours. In blue, the standard deviation of the residual for at each redshift for 1,000 different subvolumes of the same size is shown.}
    \label{fig:xhii_evo}
\end{figure*}

We then evaluate the evolution of the mean fraction of ionized hydrogen over several distinct subvolumes. \autoref{fig:xhii_evo}, top panel, shows the evolution of the ionized fraction for four different subvolumes, which each have different ionization history. Generally, we see very good agreement for $z > 8$. For early ionizing subvolumes (vol. 3 and 4), the \texttt{PINION} reionization fraction predictions are always postponed for the NP scenario. The physically constrained scenarios (PFD and LD) ionize earlier for $z > 7.5$, and later otherwise. For late ionizing subvolumes (vol. 1 and 2), as well as the global average (black line), when $z \gtrsim 7$ we observe good agreement for the LD scenario, over-prediction for the PFD scenario, and under-prediction for the NP scenario. For $z < 7$ all scenarios ionize earlier than the simulation. In general, we observe that for redshifts $z \gtrsim 7$ the LD scenario gives better results. For $z \lesssim 7$, the NP scenario is better at reproducing later ionizations, while LD remains better for earlier ionizations. In the residual plot, we observe a higher standard deviation at lower redshift ($z \lesssim 7.5$) in general. For vol. 1 and 2 the deviation peaks at $z\approx6.6$ while vol. 3 and 4 have a negative peak earlier at $z\approx7.1$. The global average shows both behaviours, with a flex point at $z\approx7$, but less residual error. This is correlated with what observed in \S~\ref{sec:network_output} for the same redshift range. In \autoref{fig:xhii_evo} bottom panel, we have the residual error of our prediction. We can see for $z \gtrsim 7$ and all subvolumes that the LD scenario follow well the simulation, while the NP scenario under-predicts with residual under \SI{2.5}{\percent} for $z \gtrsim 7.5$ while the PFD over-predicts it with less than \SI{1}{\percent} difference for $z \gtrsim 7.5$. In \autoref{tab:xhii_evo}, we show the corresponding redshift of four reionization milestones, when $\bar{x}_\mathrm{HII} = 0.3, 0.5, 0.7$ and toward reionization completion at $\bar{x}_\mathrm{HII}=0.95$, interpolated from the results in \autoref{fig:xhii_evo}.
In \S~\ref{app:massVSvol} we show a comparison between mass-averaged and volume-averaged global ionization fraction.

\subsection{Morphology of the ionization bubbles}
In the study of the EoR, of particular interest are the size and volume distributions of the ionized regions in the simulation \citep{furlanetto_growth_2004}.
There are several methods to measure these statistics \citep{giri_bubble_2018, lin_distribution_2016, friedrich_topology_2011}, but in this work we consider the mean free path method and the friend-of-friend method. The former indicates the fraction of ionized bubbles in a given spherical-averaged size range, while the latter indicates the fraction of the total ionized volume that is contained in regions of a given volume. For both methods, we employ the \texttt{Tools21cm}\footnote{\url{https://github.com/sambit-giri/tools21cm}} python package for EoR simulations analysis \citep{giri_tools21cm_2020}.

\begin{figure}
    \centering
    \includegraphics[width=0.45\textwidth]{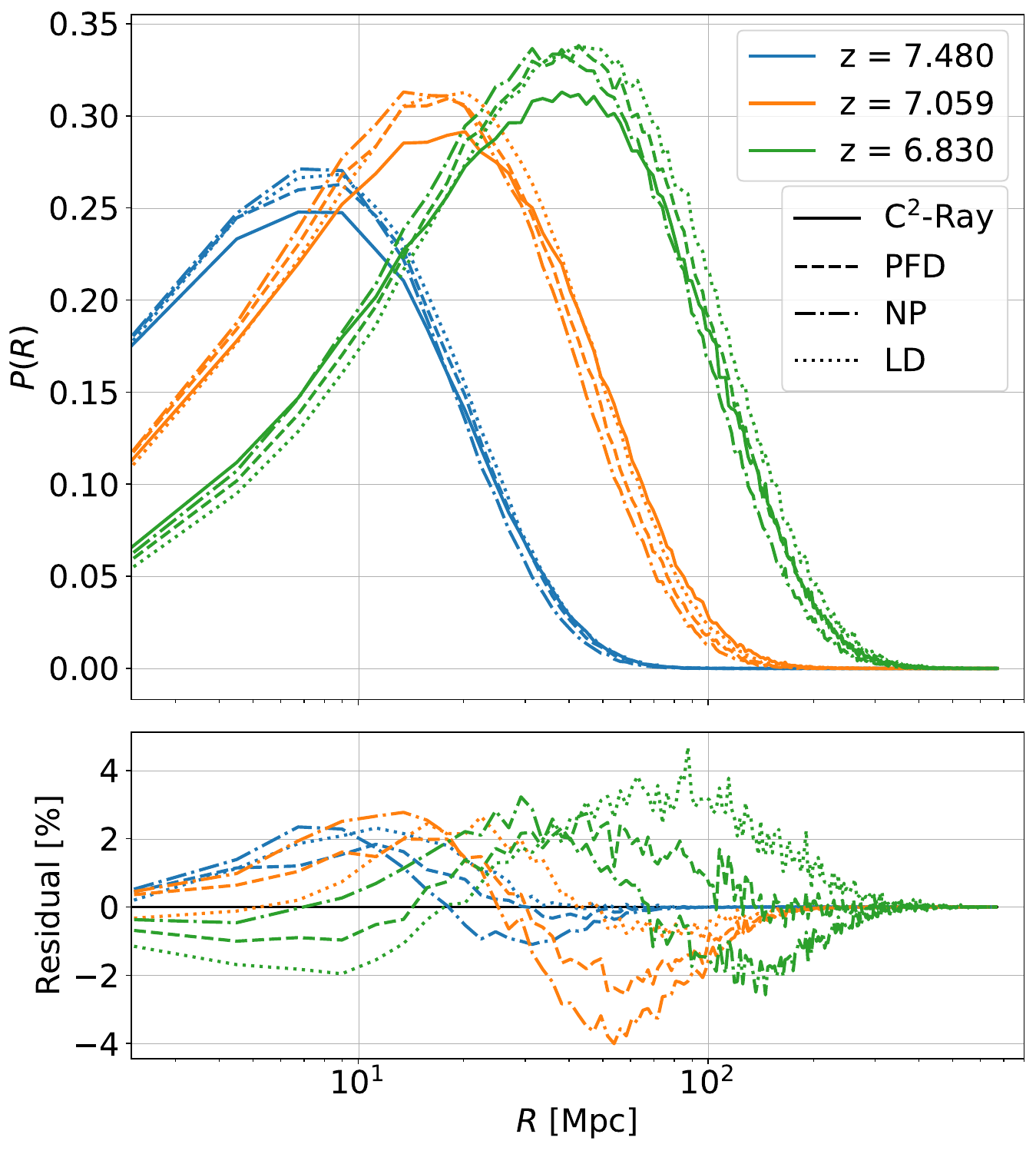}
    \vskip-4mm
    \caption{\textit{Top panel}: Study of the ionization bubbles morphology using the mean free path method. Solid lines are \texttt{C$^2$-Ray}, dashed lines are PFD training, dashdot lines are NP training and dotted lines are LD training. The lines with the same colours are picked at the same redshift. The blue line corresponds to approximately $\bar{x}_\mathrm{HII} = 0.3$, the orange to $\bar{x}_\mathrm{HII} = 0.5$ and the green to $\bar{x}_\mathrm{HII} = 0.7$. \textit{Bottom panel}: Residual plot for the above distribution. The residual is obtained by subtracting the predicted \texttt{PINION} distribution with \texttt{C$^2$-Ray}'s distribution.}
    \label{fig:morpho_mfp}
\end{figure}

\autoref{fig:morpho_mfp} shows the mean free path $R$ of the different \texttt{PINION} scenarios and the simulation plotted against the probability distribution $P(R)$. We chose to study redshifts for which $\bar{x}_\mathrm{HII} \approx 0.3, 0.5, 0.7$ which approximately corresponded to redshifts $z=7.480, 7.059, 6.830$.
All three \texttt{PINION} scenarios exhibit similar morphology. In particular, in each of them, we observe an excess around the peaks, meaning that we have a more uniform distribution of bubble size.
The fact that we observe this for all three redshift means that the universe evolves more uniformly with the \texttt{PINION} prediction than with the \texttt{C$^2$-Ray} simulation.
%
%
%

\begin{table}
    \centering
    \begin{tabular}{c|c|c|c|c|c}
        \cline{2-5}
        & \multirow{1}{*}{$z_{0.3}$} & $z_{0.5}$ & $z_{0.7}$ & $z_{0.95}$ & \\
        \hline
        \multicolumn{1}{|c|}{\multirow{4}{*}{Global}} & 7.519 & 7.108 & 6.847 & 6.510 & \multicolumn{1}{c|}{\texttt{C$^2$-Ray}} \\
        \cline{2-6}
        \multicolumn{1}{|c|}{}  & 7.545 & 7.120 & 6.891 & 6.562 & \multicolumn{1}{c|}{PFD}\\
        \cline{2-6}
        \multicolumn{1}{|c|}{}  & 7.451 & 7.066 & 6.857 & 6.512 & \multicolumn{1}{c|}{NP}\\
        \cline{2-6}
        \multicolumn{1}{|c|}{}  & 7.513 & 7.120 & 6.903 & 6.560 & \multicolumn{1}{c|}{LD}\\
        \hline\hline
        \multicolumn{1}{|c|}{\multirow{4}{*}{Vol. 1}} & 7.031 & 6.685 & 6.535 & 6.369 & \multicolumn{1}{c|}{\texttt{C$^2$-Ray}} \\
        \cline{2-6}
        \multicolumn{1}{|c|}{}  & 7.073 & 6.817 & 6.670 & 6.432 & \multicolumn{1}{c|}{PFD}\\
        \cline{2-6}
        \multicolumn{1}{|c|}{}  & 7.023 & 6.780 & 6.633 & 6.385 & \multicolumn{1}{c|}{NP}\\
        \cline{2-6}
        \multicolumn{1}{|c|}{}  & 7.059 & 6.824 & 6.676 & 6.421 & \multicolumn{1}{c|}{LD}\\
        \hline\hline
        \multicolumn{1}{|c|}{\multirow{4}{*}{Vol. 2}} & 7.299 & 6.923 & 6.711 & 6.484 & \multicolumn{1}{c|}{\texttt{C$^2$-Ray}} \\
        \cline{2-6}
        \multicolumn{1}{|c|}{}  & 7.316 & 6.974 & 6.780 & 6.533 & \multicolumn{1}{c|}{PFD}\\
        \cline{2-6}
        \multicolumn{1}{|c|}{}  & 7.233 & 6.929 & 6.744 & 6.487 & \multicolumn{1}{c|}{NP}\\
        \cline{2-6}
        \multicolumn{1}{|c|}{}  & 7.292 & 6.980 & 6.791 & 6.528 & \multicolumn{1}{c|}{LD}\\
        \hline\hline
        \multicolumn{1}{|c|}{\multirow{4}{*}{Vol. 3}} & 7.854 & 7.446 & 7.189 & 6.903 & \multicolumn{1}{c|}{\texttt{C$^2$-Ray}} \\
        \cline{2-6}
        \multicolumn{1}{|c|}{}  & 7.906 & 7.434 & 7.125 & 6.788 & \multicolumn{1}{c|}{PFD}\\
        \cline{2-6}
        \multicolumn{1}{|c|}{}  & 7.810 & 7.377 & 7.085 & 6.752 & \multicolumn{1}{c|}{NP}\\
        \cline{2-6}
        \multicolumn{1}{|c|}{}  & 7.878 & 7.428 & 7.137 & 6.792 & \multicolumn{1}{c|}{LD}\\
        \hline\hline
        \multicolumn{1}{|c|}{\multirow{4}{*}{Vol. 4}} & 8.037 & 7.592 & 7.307 & 6.965 & \multicolumn{1}{c|}{\texttt{C$^2$-Ray}} \\
        \cline{2-6}
        \multicolumn{1}{|c|}{}  & 8.095 & 7.609 & 7.282 & 6.843 & \multicolumn{1}{c|}{PFD}\\
        \cline{2-6}
        \multicolumn{1}{|c|}{}  & 7.995 & 7.548 & 7.236 & 6.806 & \multicolumn{1}{c|}{NP}\\
        \cline{2-6}
        \multicolumn{1}{|c|}{}  & 8.071 & 7.605 & 7.291 & 6.854 & \multicolumn{1}{c|}{LD}\\
        \hline
    \end{tabular}
    \vskip-1mm
    \caption{Linearly interpolated redshifts for a given volume-averaged $\bar{x}_\mathrm{HII}$. The values are derived from the results from \autoref{fig:xhii_evo}.}
    \label{tab:xhii_evo}
\end{table}

For redshifts $z=7.480$ (blue) and $z=7.059$ (orange), \texttt{PINION} predicts an excess of smaller scale bubbles below the median, and a lack of larger scale bubbles above the median compared to \texttt{C$^2$-Ray}.
For redshift $z=7.480$ (blue), only the NP scenario lacks large scale bubbles ($R\gtrsim \SI{20}{\mega\parsec}$), which means that the NP scenario produces smaller bubbles on average than the PFD and LD scenarios.
For redshift $z=7.059$ (orange), however, we observe that all three scenarios have a turning point above which the prediction lacks large scale bubbles.
For large and small scale bubbles, the LD scenario gives a more accurate distribution, but around each peak, the PFD scenario tend to be closer to the distribution in the simulation.

For redshift $z=6.830$ (green), we observe a deficit of small-scale bubbles for all scenarios.
In general, we observe that the predicted distributions are shifted to the larger scale, meaning that the bubbles are larger on average than in the simulation.

\begin{figure}
    \centering
    \includegraphics[width=.45\textwidth]{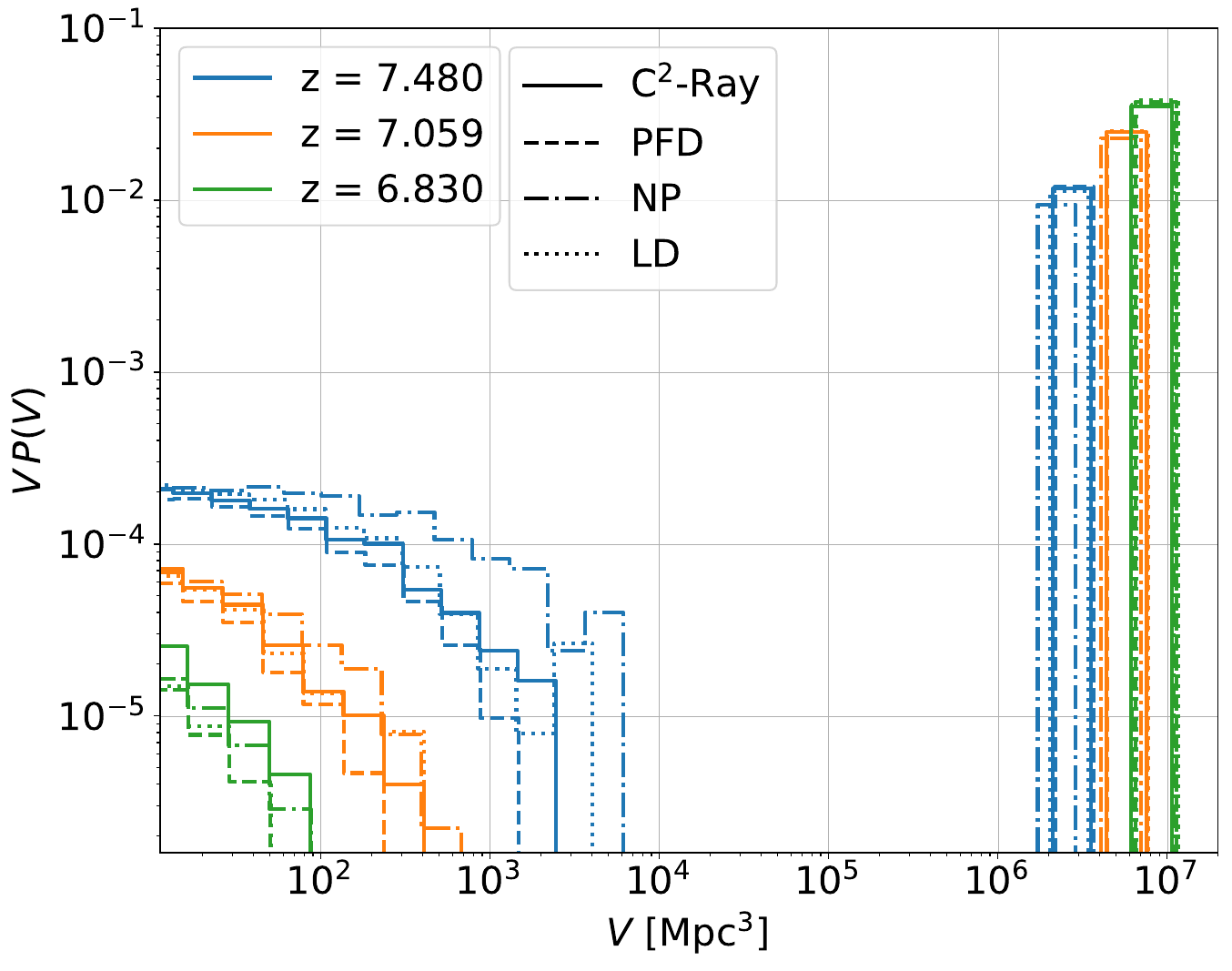}
    \vskip-4mm
    \caption{Study of the ionization bubbles volume distribution using the friend-of-friend method. Solid lines are \texttt{C$^2$-Ray}, dashed lines are PFD training, dashdot lines are NP training and dotted lines are LD training. Both lines are picked at the same redshift. The blue line corresponds to approximately $\bar{x}_\mathrm{HII} = 0.3$, the orange to $\bar{x}_\mathrm{HII} = 0.5$ and the green to $\bar{x}_\mathrm{HII} = 0.7$.}
    \label{fig:morpho_fof}
\end{figure}

We also study the volume distribution with the friend-of-friend method. Contrary to the mean free path, which measures the size of the bubbles, the friend-of-friend algorithm evaluates the volume distribution of the bubbles using an island-finding algorithm. Pixels are recursively grouped into islands based on a threshold criterion, in our case of $x_\mathrm{HII} > 0.5$. The average volume of each island $V$ is then plotted against the cumulative probability distribution $V\,P(V)$. \autoref{fig:morpho_fof} shows the result for the friend-of-friend analysis. First, for redshift $z=7.480$ (blue), one can see that both the PFD and LD scenarios are able to accurately reconstruct the bubble volume distribution and the \textit{percolation cluster}\footnote{The volume formed by the overlapping bubbles.}. On the other hand, the NP scenario tends to over-predict the number of non-overlapping bubbles, as one can observe an over-prediction of the bubbles distribution and an underestimation of the total volume of the percolation cluster. For redshifts $z=7.059$ (orange) and $z=6.830$ (green) all three scenarios are able to reproduce the volume distribution of both the bubbles and the percolation cluster. In general, we observe that the volume size distribution is well reproduced by \texttt{PINION} regardless of the scenario, aside from NP at higher redshifts. 

\subsection{HI dimensionless power spectrum} \label{sec:power_spectra}
\begin{figure}
    \centering
    \includegraphics[width=.5\textwidth]{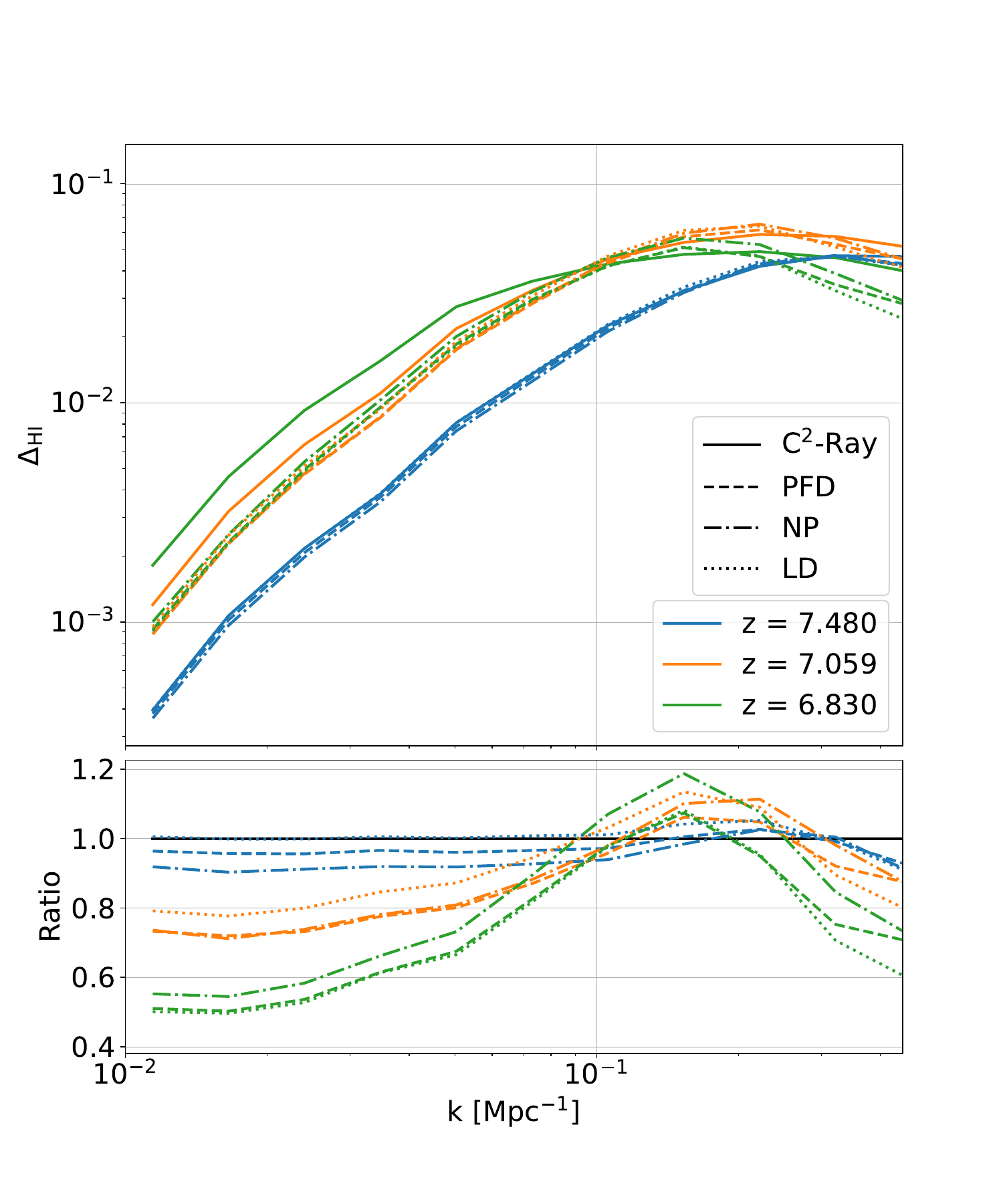}
    \vskip-10mm
    \caption{\textit{Top panel}: Dimensionless power spectrum of $x_\mathrm{HI}$. Solid lines are \texttt{C$^2$-Ray}, dashed lines are PFD training, dash-dot lines are NP training and dotted lines are LD training. These lines are picked at the same redshifts. The blue line corresponds to approximately $\bar{x}_\mathrm{HI} = 0.7$, the orange to $\bar{x}_\mathrm{HI} = 0.5$ and the green to $\bar{x}_\mathrm{HI} = 0.3$. \textit{Bottom panel}: Ratio plot for the above dimensionless power spectrum. The ratio is obtained by dividing the predicted power spectrum with \texttt{C$^2$-Ray}'s power spectrum.}
    \label{fig:ps}
\end{figure}

We also evaluate the dimensionless power spectrum of the neutral hydrogen field $\Delta_\mathrm{HI} = P(k) \frac{k^3}{2\pi^2}$ for both \texttt{C$^2$-Ray} and \texttt{PINION}, shown in \autoref{fig:ps}. We observe almost perfect agreement in large and small scale structure for $z=7.480$ (blue). However, as ionization progresses, we begin to see tension between \texttt{PINION} and \texttt{C$^2$-Ray}. At larger scales, $k < 10^{-1}\,\rm{Mpc^{-1}}$, the power spectrum is underestimated by all \texttt{PINION} predictions at all redshifts, meaning a statistically more ionized field then the ground truth. We observe that at these scales the discrepancy between the \texttt{C$^2$-Ray} simulation and the \texttt{PINION} prediction increases with decreasing redshift, from almost perfect agreement at $z=7.480$ (blue) to up to a factor of 2 difference for  $z=6.830$ (green).
\vfill

\subsection{Reionization redshift}
\begin{figure*}
    \centering
    \makebox[\textwidth][c]{\includegraphics[width=\textwidth]{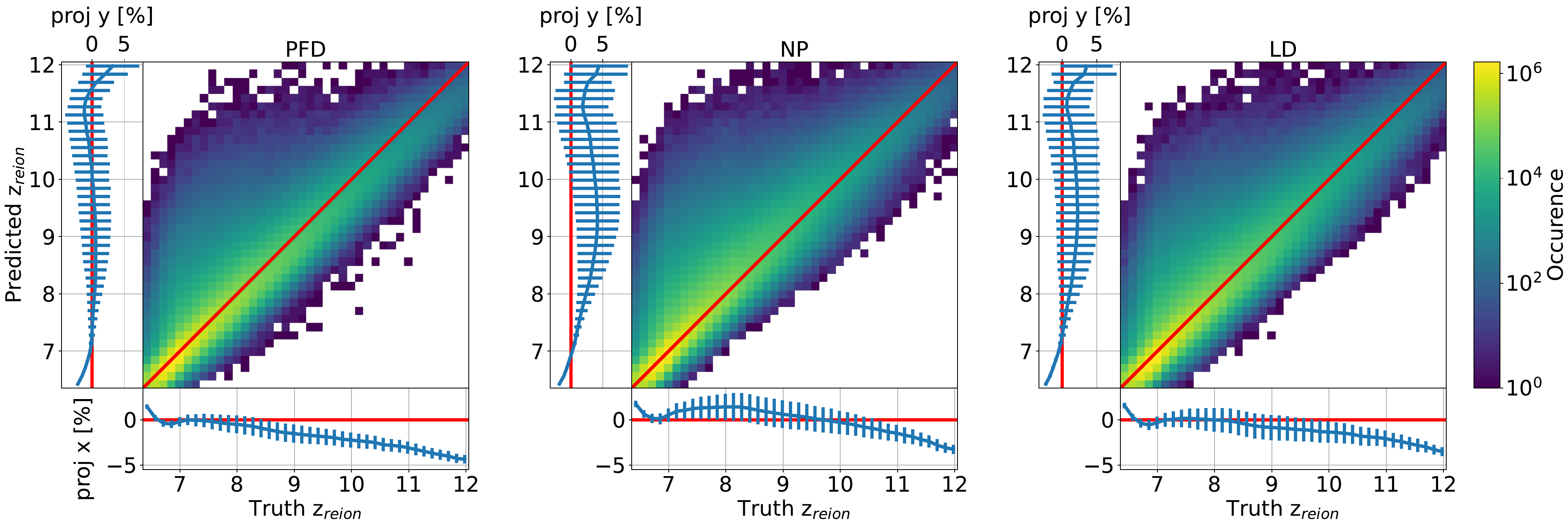}}
    \caption{Comparison between the \texttt{PINION} prediction and the \texttt{C$^2$-Ray} prediction of the redshift of reionisation $z_\text{reion}$. For each training scenario, the \textit{central panel} shows the 2D histogram of the predicted and truth redshift at which a region is considered ionized. A region is considered ionized when it reaches $x_{\mbox{HII}} \geq 0.5$. We linearly interpolated the redshift to have the value at the threshold to avoid tessellation effects. The red line shows emphasis the 1:1 relationship, which would correspond to a perfect result.
    The \textit{vertical plots} show the y projection of the histogram, with the expected value and standard deviation of the residual in blue. The red line emphasis the perfect prediction, which is a residual of 0. The value of the residual is divided by the central redshift of the bins of interest. The \textit{horizontal plots} show the x projection. The axis and the colors gradient are shared between scenarios to improve readability.}
    \label{fig:reionization_redshift_map}
\end{figure*}

From the predicted reionsation maps, we can calculate the reionization redshift $z_\text{reion}$. Defined as the redshift at which a pixel becomes ionized fraction is $x_{\mbox{HII}} \geq 0.5$. In \autoref{fig:reionization_redshift_map} we compare the true and reconstructed $z_\text{reion}$ for our three training scenarios.

At first glance, all three histograms look very similar, but some differences can be observed in the x projection. We observe a larger residual at higher redshifts, with \SI{-4.3(5)}{\percent}, \SI{-3.1(6)}{\percent}, \SI{-3.3(6)}{\percent} for $z=\num{11.834}$ in the PFD, NP and LD scenarios respectively. As shown in \S\ref{sec:power_spectra}, this indicates that \texttt{PINION} ionizes earlier than \texttt{C$^2$-Ray} at high redshift. However, this difference reduces with redshift and in the particular case of NP, will have a positive peak of \SI{1.4\pm 1.5}{\percent} at $z=\num{8.133}$ which this time indicates that at this ionization redshift, \texttt{PINION} tends to ionize later compared to \texttt{C$^2$-Ray}. For the y projection, we observe a large positive bias in both NP and LD with \SI{4.1 \pm 3.5}{\percent} for NP and \SI{2.2 \pm 3.1}{\percent} at $z=\num{9.414}$ for LD.
Near the end of the EoR, $z \leq 7$, we observe a sudden change in the residuals of every scenario. We believe this results from small variations on the borders of  neutral islands, which are difficult for the network to predict precisely.

We can also use \autoref{fig:reionization_redshift_map} to compare \texttt{PINION} to 
\citep{chardin_deep_2019}, a network which predicts the 3D $z_\text{reion}$ map. 
PINION shows much better agreement at early and mid redshift, likely due to the addition of the physics constraint. Additionally, because PINION outputs $x_{\mbox{HII}}$, while PINION can reproduce \citep{chardin_deep_2019}, the inverse is not possible.


\subsection{Interpretation}\label{sec:interp}
We observe a few consistent differences between  \texttt{C$^2$-Ray} and the \texttt{PINION} predictions in our analysis. \texttt{PINION} predictions are more uniform, having more uniform bubble sizing and less variety in mean ionization fraction. The NP scenario ionizes later at high redshift, whereas the physics-constrained scenarios PDF and LD ionize sooner. For redshifts with rapidly changing ionization fraction, at $z \lesssim 7$, there is tension between the \texttt{PINION} predictions and \texttt{C$^2$-Ray} at small and particularly large scales. There are several possible explanations for these systematic errors.

\subsubsection*{Data bias}
The \texttt{C$^2$-Ray} simulation used to train \texttt{PINION} is an imbalanced dataset depending on the redshift. For instance, the vast majority of pixels at high redshift, early reionisation history, are neutral rather than ionized. As such, it is possible that this biases the network predictions towards more neutral predictions. Any such bias would be most evident in the NP scenario, and we see for $z > 7$ in \autoref{fig:xhii_evo} that the NP scenario always predicts delayed ionization, thus more neutral volumes compared to truth. Additionally, in comparison to the PFD and LD scenarios, the NP has larger neutral islands and smaller ionized bubbles.


\subsubsection*{Finite-difference instabilities}
Instabilities coming from the finite-difference approximation used to calculate the derivative can be observed as a direct issue for the predictions. Although \texttt{PINION} uses a Runge-Kutta 4 scheme for differentiation, which has an error of the order $\mathcal{O}(h^5)$, the time steps between each snapshot are large compared to what is required by the scheme. We expect the tension introduced by this bias to affect the LD and PFD scenarios, and we indeed observe that the NP scenario has the best agreement with \texttt{C$^2$-Ray} for $z < 7$. The finite-difference instability issue is observed, as shown in \autoref{fig:rk4_v_c2r}, but the maximal observed difference lies within \SI{2}{\percent}. This is lower than the residual that we observe in \autoref{fig:xhii_evo}, where a standard deviation within \SI{12}{\percent} is observed for $z<7$. While instabilities of the derivative evaluation do introduce a slight bias, this does not account for the majority of the differences observed between \texttt{C$^2$-Ray} and \texttt{PINION}.

\subsubsection*{Inaccuracies from smoothed source distribution field}
The last main issue encountered in this project is the use of the smoothed source distribution field.
As described in \S~\ref{sec:prop_mask}, it is a way to provide additional information to the network, given that the two outputs from the N-Body simulation do not provide enough information about the growing structure of the ionized regions.
However, as the smoothed field is generated by convolving uniform spheres with the input fields, it is much more uniform in structure compared to the behaviour of the ionization front and therefore the photoionization rate.
The morphology of the ionization rate map and the smoothed source distribution field are compared in appendix \S~\ref{app:valididy}, and in \S~\ref{app:f_mfp} we test the influence of the smoothed source distribution field by adjusting the scale length of \autoref{eq:mfp_photons} on the reionization history.